# Agile Software Development Methods:
# Review and Analysis

## Authors: Pekka Abrahamsson, Outi Salo, Jussi Ronkainen and Juhani Warsta

# Contents

# 1. Introduction

The field of software development is not shy of introducing new methodologies. Indeed, in the last 25 years, a large number of different approaches to software development have been introduced, of which only few have survived to be used today. A recent study (Nandhakumar and Avison 1999) argues that traditional information systems[1] (IS) development methodologies "are treated primarily as a necessary fiction to present an image of control or to provide a symbolic status." The same study further claims that these methodologies are too mechanistic to be used in detail. Parnas and Clements (1986) have made similar arguments early on. Truex et al. (2000) take an extreme position and state that it is possible that traditional methods are "merely unattainable ideals and hypothetical 'straw men' that provide normative guidance to utopian development situations". As a result, industrial software developers have become skeptical about "new" solutions that are difficult to grasp and thus remain not used (Wiegers 1998). This is the background for the emergence of agile software development methods.

While no agreement on what the concept of "agile" actually refers to exists, it has generated a lot of interest among practitioners and lately also in the academia. The introduction of the extreme programming method (Better known as the XP, Beck 1999a; Beck 1999b) has been widely acknowledged as the starting point for the various agile software development approaches. There are also a number of other methods either invented or rediscovered since then that appear to belong to the same family of methodologies. Such methods or methodologies are, e.g., Crystal Methods (Cockburn 2000), Feature-Driven Development (Palmer and Felsing 2002), and Adaptive Software Development (Highsmith 2000). As a sign of the increased interest, the Cutter IT Journal has recently dedicated three full issues to the treatment of light methodologies, and the participation of at least two major international conferences has had to be limited due to a high number of attendees.

[1] Software engineering (SE) differs from the field of IS predominantly in the sense that the IS community takes into account the social and organizational aspects (e.g., Dhillon 1997; Baskerville 1998). Moreover, SE traditionally focuses on practical means of developing software (Sommerville 1996). However, for the purposes of this publication such a distinction is not necessary. Thus, IS literature concerning the actual use of different methods is considered relevant.

While little is known about the actual payoff of the investment made into process technologies (Glass 1999), even less is known about how much an organization will benefit from the use of agile software development methodologies. The initial experience reports from industry are predominantly positive (e.g., Anderson et al. 1998; Williams et al. 2000; Grenning 2001). Hard numbers, however, are difficult to obtain at this stage.

Despite the high interest in the subject, no clear agreement has been achieved on how to distinguish agile software development from more traditional approaches. The boundaries – if such exist – have thus not been clearly established. However, it has been shown that certain methods are not necessarily suitable for all individuals (Naur 1993) or settings (Baskerville et al. 1992). For this reason, e.g. Humphrey (1995) calls for the development of a personal process for each software developer. Despite these findings little emphasis has been placed on analyzing for which situations agile methods are more suitable than others. To our knowledge, no systematic review of agile development methodologies has been done yet. As a result of this, currently, there are no procedures available for the practitioner for choosing the method bringing the greatest benefit for the given circumstances.

Our goal, therefore, is to begin filling this gap by systematically reviewing the existing literature on agile software development methodologies. This publication has thus three purposes. Firstly, as a result of this synthesizing analysis a definition and a classification of agile approaches is proposed. Secondly, an analysis of the proposed approaches against the defined criterion is provided, and thirdly, agile development methods introduced are compared in order to highlight their similarities and differences.

This work is organized as five sections. In the following section, a definition of the agile software development method as used in the context of this publication is provided. The third section reviews most of the existing agile software development methods, which are subsequently compared, discussed and summarized in section four. In the sixth section, the publication is concluded with final remarks.

# 2. Agile overview, definitions and characterizations

The purpose of this section is to characterize the meanings that are currently associated with the concept of "agile", and to provide a definition of the agile software development method as used in the context of this publication.

## 2.1. Background

Agile – denoting "the quality of being agile; readiness for motion; nimbleness, activity, dexterity in motion[2]" – software development methods are attempting to offer once again an answer to the eager business community asking for lighter weight along with faster and nimbler software development processes. This is especially the case with the rapidly growing and volatile Internet software industry as well as for the emerging mobile application environment. The new agile methods have evoked substantial amount of literature and debates, see e.g. the following feature articles in the Cutter IT Journal[3]: The Great Methodologies Debate: Part 1 and Part 2. However, academic research on the subject is still scarce, as most of existing publications are written by practitioners or consultants.

Truex et al. (2000) question if the information systems development is in practice actually executed following the rules and guidelines defined by the numerous software development methods available. The differences of privileged and marginalized methodological information systems development processes as proposed by the authors are presented in Table 1.

*Table 1. Some differences of the privileged and marginalized methodological ISD processes*

[2] dictionary.oed.com, (2.5.2002)

[3] Cutter IT Journal, November (Vol. 14, No. 11), December 2001 (Vol. 14, No. 12) and January 2002 (Vol. 15, No. 1)

| **Privileged methodological text** | **Marginalized methodological text** |
|---|---|
| Information systems development is: | |
| A managed, controlled process | Random, opportunistic process driven by accident |
| A linear, sequential process | Processes are simultaneous, overlapping and there are gaps in between |
| A replicable, universal process | Occurs in completely unique and idiographic forms |
| A rational, determined, and goal-driven process | Negotiated, compromised and capricious |

This comparison gives an interesting and an enlightening perspective and provides some background for analyzing the agile software development methods (a.k.a. light methods). Privileged method projects use commonly accepted processual (also known as plan-driven) software development methods, while marginalized methods have much in common with the novel agile software development methods, which are discussed in more depth below.

McCauley (2001) argues that the underlying philosophy of process-oriented methods is that the requirements of the software project are completely locked in and frozen before the design and software development commences. As this approach is not always feasible there is a need also for flexible, adaptable and agile methods, which allow the developers to make late changes in the specifications.

## 2.2. Overview and definitions

The "Agile Movement" in software industry saw the light of day with the Agile Software Development Manifesto[4] published by a group of software practitioners and consultants in 2001 (Beck et al. 2001; Cockburn 2002a). The focal values honored by the agilists are presented in the following:

- **Individuals and interactions** over processes and tools
- **Working software** over comprehensive documentation
- **Customer collaboration** over contract negotiation
- **Responding to change** over following a plan

These central values that the agile community adheres to are:

*First*, the agile movement emphasizes the relationship and communality of software developers and the human role reflected in the contracts, as opposed to institutionalized processes and development tools. In the existing agile practices, this manifests itself in close team relationships, close working environment arrangements, and other procedures boosting team spirit.

*Second*, the vital objective of the software team is to continuously turn out tested working software. New releases are produced at frequent intervals, in some approaches even hourly or daily, but more usually bi-monthly or monthly. The developers are urged to keep the code simple, straightforward, and technically as advanced as possible, thus lessening the documentation burden to an appropriate level.

*Third*, the relationship and cooperation between the developers and the clients is given the preference over strict contracts, although the importance of well drafted contracts does grow at the same pace as the size of the software project. The negotiation process itself should be seen as a means of achieving and maintaining a viable relationship. From a business point of view, agile development is focused on delivering business value immediately as the project starts, thus reducing the risks of non-fulfillment regarding the contract.

[4] agilemanifesto.org and www.agilealliance.org, (1.5.2002)

*Fourth*, the development group, comprising both software developers and customers representatives, should be well-informed, competent and authorized to consider possible adjustment needs emerging during the development process life-cycle. This means that the participants are prepared to make changes and that also the existing contracts are formed with tools that support and allow these enhancements to be made.

According to Highsmith and Cockburn (2001, p. 122), "what is new about agile methods is not the practices they use, but their recognition of people as the primary drivers of project success, coupled with an intense focus on effectiveness and maneuverability. This yields a new combination of values and principles that define an *agile* world view." Boehm (2002) illustrates the spectrum of different planning methods with Figure 1, in which hackers are placed at one end and the so called inch-pebble ironbound contractual approach at the opposite end:

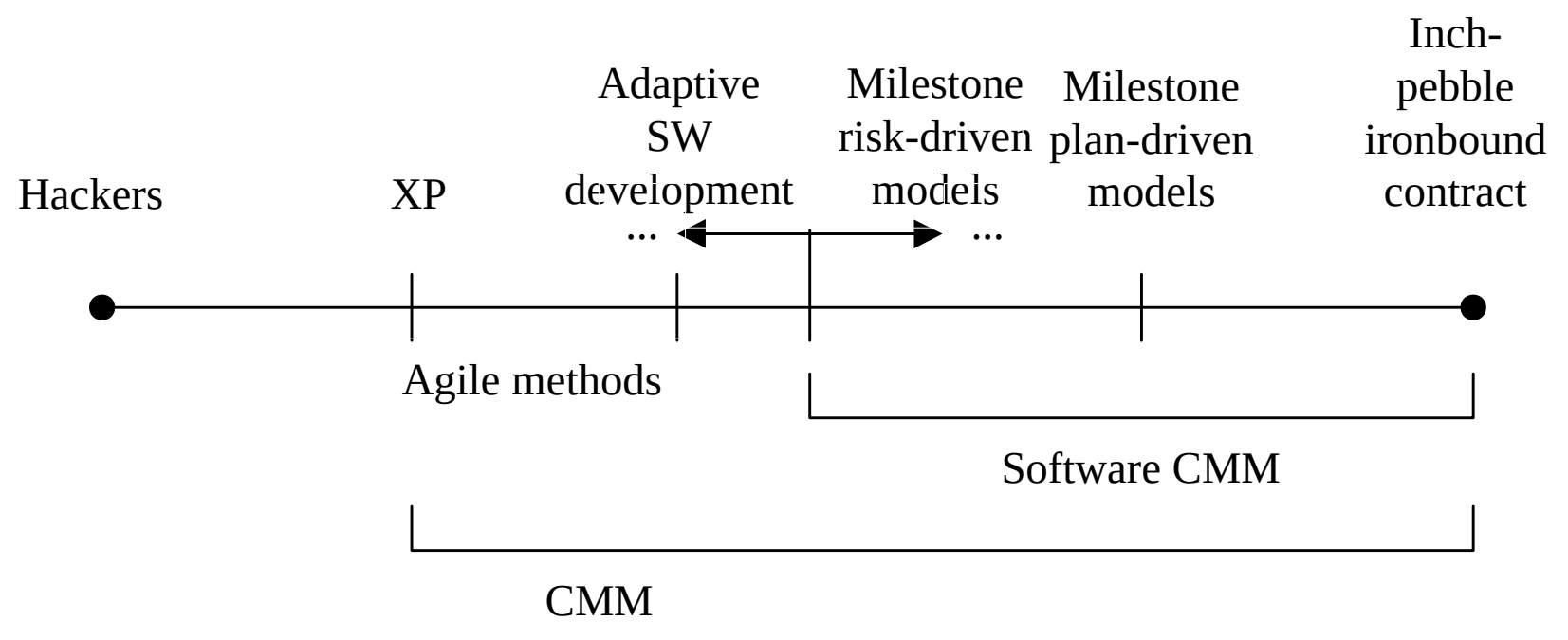

*Figure 1. The planning spectrum (Boehm 2002, p. 65)*

Hawrysh and Ruprecht (2000) state that a single methodology can not work for the whole spectrum of different projects, but instead the project management should identify the specific nature of the project at hand and then select the best applicable development methodology. To stress the point further, according to McCauley (2001) there is a need for both agile and process-oriented methods, as there is no one-size-fits-all software development model that suits all imaginable purposes. This opinion is shared by several experts in the field (Glass 2001).

Cockburn (2002a, p. xxii) defines the core of agile software development methods as the use of light-but-sufficient rules of project behavior and the use of human- and communication-oriented rules. The agile process is both light and sufficient. Lightness is a means of remaining maneuverable. Sufficiency is a matter of staying in the game (Cockburn 2002a). He proposes the following "sweet spots" the presence of which in software development work enhances the prospects for a successful project outcome:

- Two to eight people in one room
  - Communication and community
- Onsite usage experts
  - Short and continuous feedback cycles
- Short increments
  - One to three months, allows quick testing and repairing
- Fully automated regression tests
  - Unit and functional tests stabilize code and allow continuous improvement
- Experienced developers
  - Experience speeds up the development time from 2 to 10 times compared to slower team members.

## 2.3. Characterization

Miller (2001) gives the following characteristics to agile software processes from the fast delivery point of view, which allow shortening the life-cycle of projects:

1. Modularity on development process level
2. Iterative with short cycles enabling fast verifications and corrections
3. Time-bound with iteration cycles from one to six weeks
4. Parsimony in development process removes all unnecessary activities
5. Adaptive with possible emergent new risks
6. Incremental process approach that allows functioning application building in small steps
7. Convergent (and incremental) approach minimizes the risks
8. People-oriented, i.e. agile processes favor people over processes and technology
9. Collaborative and communicative working style

Favaro (2002) discusses the emergence of strategies for confronting a vague process showing changing requirements. He proposes the iterative development paradigm as the common denominator of agile processes. Requirements may be introduced, modified or removed in successive iterations. Once more, this approach featuring changing requirements and delayed implementation calls for new contractual conventions. These are, e.g., transition from point-view contracts (nailing down the requirements up front) towards process-view contracts, and also the consideration of the anticipation of legal aspects in relationship evolvement (Pöyhönen 2000). These theoretical legal aspects, however, are still in their beginning, not to mention the concrete capitalization of these new contractual phenomena. Also the framework contracts efficiently used together with relevant project or work order agreements reflect the ongoing development in software business, which inherently supports this kind of agile software development (Warsta 2001).

Highsmith and Cockburn (2001) report that the changing environment in software business also affects the software development processes itself. According to the authors, to satisfy the customers at the time of delivery has taken precedence over satisfying the customer at the moment of the project initiation. This calls for procedures not so much dealing with how to stop change early in a project, but how to better handle inevitable changes throughout its life cycle. It is further claimed that agile methods are designed to:

- produce the first delivery in weeks, to achieve an early win and rapid feedback

- invent simple solutions, so there is less to change and making those changes is easier

- improve design quality continually, making the next story less costly to implement, and

- test constantly, for earlier, less expensive, defect detection.

The basic principles of agile methods comprise an unforgiving honesty of working code, effectiveness of people working together with goodwill, and focus on teamwork. A set of commonsense approaches emerging from agile software development processes have been suggested by Ambler (2002b) as follows:

- people matter

- less documentation is possible

- communication is a critical issue

- modeling tools are not as useful as usually thought

- big up-front design is not required.

Boehm (2002) analyses the agile and process-oriented methodologies or plan-driven as he calls them. Table 2 shows how the Open Source Software (OSS) paradigm places itself between the agile and plan-driven methods. The OSS is still fairly new in business environment and a number of interesting research questions remain to be analyzed and answered. Thus the OSS approach can be seen as one variant of the multifaceted agile methods.

*Table 2. Home ground for agile and plan-driven methods (Boehm 2002), augmented with open source software column.*

| **Home-ground area** | **Agile methods** | **Open source software** | **Plan-driven methods** |
|---|---|---|---|
| Developers | Agile, knowledgeable, collocated, and collaborative | Geographically distributed, collaborative, knowledgeable and agile teams | Plan-oriented; adequate skills; access to external knowledge |
| Customers | Dedicated, knowledgeable, collocated, collaborative, representative, and empowered | Dedicated , knowledgeable, collaborative, and empowered | Access to knowledgeable, collaborative, representative, and empowered customers |
| Requirements | Largely emergent; rapid change | Largely emergent; rapid change, commonly owned, continually evolving – “never” finalized | Knowable early; largely stable |
| Architecture | Designed for current | Open, designed for current requirements | Designed for current and foreseeable requirements |

| **Home-ground area** | **Agile methods** | **Open source software** | **Plan-driven methods** |
|---|---|---|---|
| | requirements | | |
| Refactoring | Inexpensive | Inexpensive | Expensive |
| Size | Smaller teams and products | Larger dispersed teams and smaller products | Larger teams and products |
| Primary objective | Rapid value | Challenging problem | High assurance |

## 2.4. Summary

The focal aspects of light and agile methods are simplicity and speed. In development work, accordingly, the development group concentrates only on the functions needed at first hand, delivering them fast, collecting feedback and reacting to received information. Based on the above discussion, a definition is proposed for the agile software development approach, and later used in this publication.

What makes a development method an agile one? This is the case when software development is **incremental** (small software releases, with rapid cycles), **cooperative** (customer and developers working constantly together with close communication), **straightforward** (the method itself is easy to learn and to modify, well documented), and **adaptive** (able to make last moment changes).

# 3. Existing agile methods

In this chapter, the current state of agile software development methods are reviewed. The selection of methods is based on the definition proposed in 2.2. As a result the following methods are included in this analysis: Extreme programming (Beck 1999b), scrum (Schwaber 1995; Schwaber and Beedle 2002), crystal family of methodologies (Cockburn 2002a), feature driven development (Palmer and Felsing 2002), the rational unified process (Kruchten 1996; Kruchten 2000), dynamic systems development method (Stapleton 1997), adaptive software development (Highsmith 2000) and open source development (e.g., O'Reilly 1999).

Methods will be introduced and reviewed systematically by using a defined structure where process, roles and responsibilities, practices, adoption and experiences, scope of use and current research regarding each method are identified. Process refers to the description of phases in the product life-cycle through which the software is being produced. Roles and responsibilities refer to the allocation of specific roles through which the software production in a development team is carried out. Practices are concrete activities and workproducts that a method defines to be used in the process. Adoption and experiences refer primarily to existing experience reports regarding the use of method in practice and method developers’ considerations how the method should be introduced in an organization. Scope of use identifies the limitations regarding each method, i.e. if such has been documented. Finally, the current research and publications are surveyed in order to get an overview of the scientific and practical status of the method.

Agile modeling (Ambler 2002a) and pragmatic programming (Hunt 2000) are introduced briefly in the last section called “Other agile methods”. This is due to a fact that they are not methods *per se* but have gained considerable attention in the agile community and thus need to be addressed.

## 3.1. Extreme Programming

Extreme Programming (XP) has evolved from the problems caused by the long development cycles of traditional development models (Beck 1999a). It first started as 'simply an opportunity to get the job done' (Haungs 2001) with practices that had been found effective in software development processes during the preceding decades (Beck 2000). After a number of successful trials in practice (Anderson et al. 1998), the XP methodology was "theorized" on the key principles and practices used (Beck 2000). Even though the individual practices of XP are not new as such, in XP they have been collected and lined up to function with each other in a novel way thus forming a new methodology for software development. The term 'extreme' comes from taking these commonsense principles and practices to extreme levels (Beck 2000).

### 3.1.1. Process

The life cycle of XP consists of five phases: Exploration, Planning, Iterations to Release, Productionizing, Maintenance and Death (Figure 2).

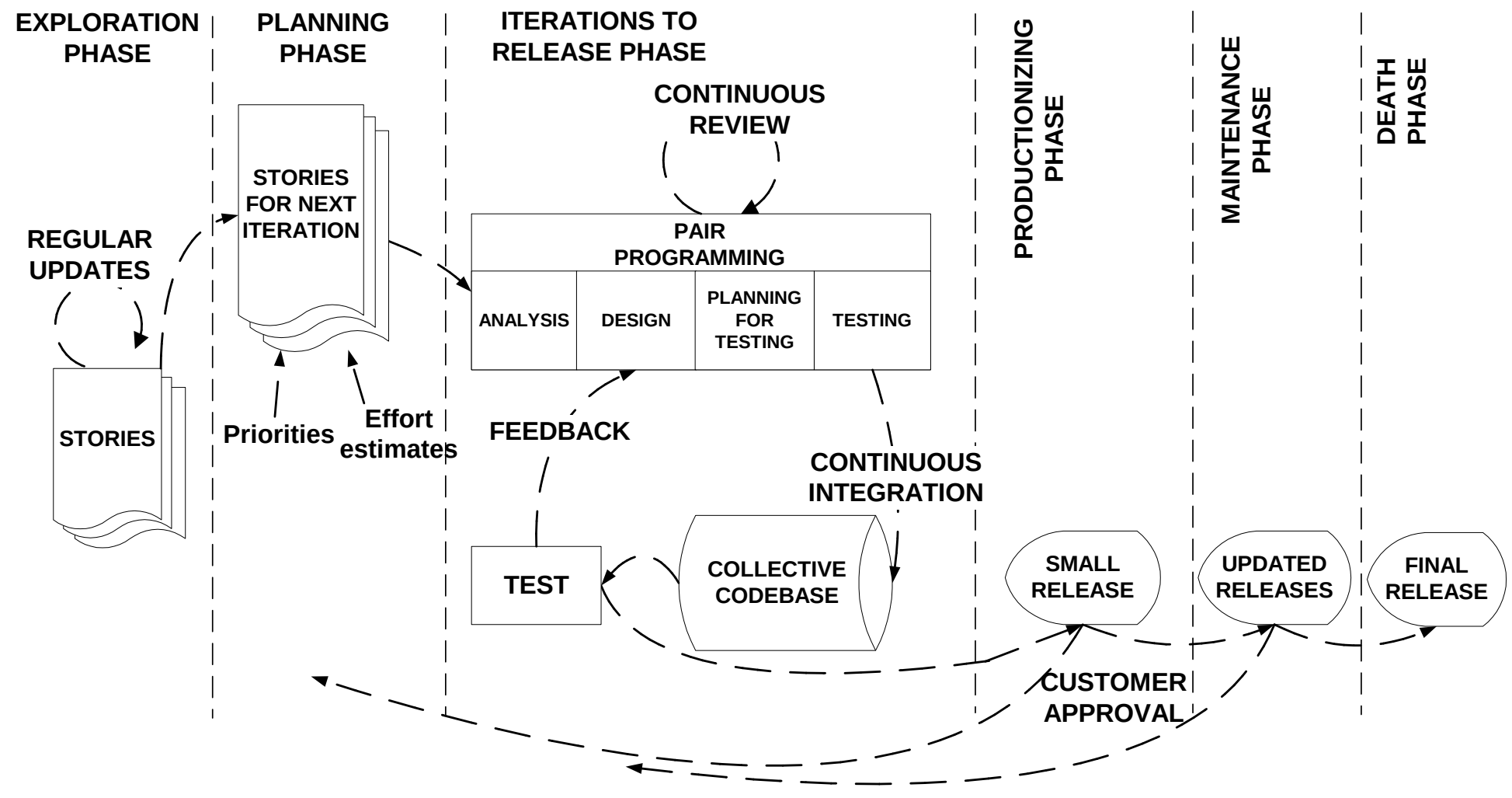

*Figure 2. Life cycle of the XP process*

In the following, these phases are introduced according to Beck's (2000) description:

In the **Exploration phase,** the customers write out the story cards that they wish to be included in the first release. Each story card describes a feature to be added into the program. At the same time the project team familiarize themselves with the tools, technology and practices they will be using in the project. The technology to be used will be tested and the architecture possibilities for the system are explored by building a prototype of the system. The exploration phase takes between a few weeks to a few months, depending largely on how familiar the technology is to the programmers.

The **Planning phase** sets the priority order for the stories and an agreement of the contents of the first small release is made. The programmers first estimate how much effort each story requires and the schedule is the agreed upon. The time span of the schedule of the first release does not normally exceed two months. The planning phase itself takes a couple of days.

The **Iterations to release** phase includes several iterations of the systems before the first release. The schedule set in the planning stage is broken down to a number of iterations that will each take one to four weeks to implement. The first iteration creates a system with the architecture of the whole system. This is achieved by selecting the stories that will enforce building the structure for the whole system. The customer decides the stories to be selected for each iteration. The functional tests created by the customer are run at the end of every iteration. At the end of the last iteration the system is ready for production.

The **Productionizing phase** requires extra testing and checking of the performance of the system before the system can be released to the customer. At this phase, new changes may still be found and the decision has to be made if they are included in the current release. During this phase, the iterations may need to be quickened from three weeks to one week. The postponed ideas and suggestions are documented for later implementation during, e.g., the maintenance phase.

After the first release is productionized for customer use, the XP project must both keep the system in the production running while also producing new iterations. In order to do this, the **Maintenance phase** requires an effort also for customer support tasks. Thus, the development velocity may decelerate after the system is in production. The maintenance phase may require incorporating new people into the team and changing the team structure.

The **Death phase** is near when the customer does no longer have any stories to be implemented. This requires that the system satisfies customer needs also in other respects (e.g., concerning performance and reliability). This is the time in the XP process when the necessary documentation of the system is finally written as no more changes to the architecture, design or code are made. Death may also occur if the system is not delivering the desired outcomes, or if it becomes too expensive for further development.

### 3.1.2. Roles and responsibilities

There are different roles in XP for different tasks and purposes during the process and its practices. In the following, these roles are presented according to Beck (2000).

**Programmer**
Programmers write tests and keep the program code as simple and definite as possible. The first issue making XP successful is to communicate and coordinate with other programmers and team members.

**Customer**
The customer writes the stories and functional tests, and decides when each requirement is satisfied. The customer sets the implementation priority for the requirements.

**Tester**
Testers help the customer write functional tests. They run functional tests regularly, broadcast test results and maintain testing tools.

**Tracker**
Tracker gives feedback in XP. He traces the estimates made by the team (e.g. effort estimates) and gives feedback on how accurate they are in order to improve future estimations. He also traces the progress of each iteration and evaluates whether the goal is reachable within the given resource and time constraints or if any changes are needed in the process.

**Coach**

Coach is the person responsible for the process as a whole. A sound understanding of XP is important in this role enabling the coach to guide the other team members in following the process.

**Consultant**

Consultant is an external member possessing the specific technical knowledge needed. The consultant guides the team in solving their specific problems.

**Manager (Big Boss)**

Manager makes the decisions. In order to be able to do this, he communicates with the project team to determine the current situation, and to distinguish any difficulties or deficiencies in the process.

### 3.1.3. Practices

XP is a collection of ideas and practices drawn from already existing methodologies (Beck 1999a). The decision making structure, as presented in Figure 3, in which the customer makes business decisions while programmers decide on technical issues is derived from the ideas of Alexander (1979). The rapid type evolution in XP has its roots in the ideas behind Scrum[5] (Takeuchi and Nonaka 1986) and the pattern language[6] of Cunningham (1996). The XP idea of scheduling projects based on customer stories is drawn from use cases[7]

---

[5] Scrum can also be regarded as an agile method. It will be introduced in section 3.2.

[6] A pattern is a named piece of insight that reveals the essence of a proven solution to a certain recurring problem within a certain context. Pattern language is a collective of patterns which work together to resolve a complex problem into an orderly solution according to a pre-defined goal. (Alexander 1979).

[7] Use cases are used for capturing the high level user-functional requirements of a system (Jacobsen 1994).

(Jacobsen 1994) and the evolutional delivery is adopted from Gilb (Gilb 1988). Also the spiral model, the initial response to the waterfall model, has had an influence on the XP method. The metaphors of XP originate from the works of Lakoff and Johnson (Lakoff and Johnson 1998), and Coyne (Coyne 1995). Finally, the physical working environment for programmers are adopted from (Coplien 1998) and (DeMarco and Lister 1999).

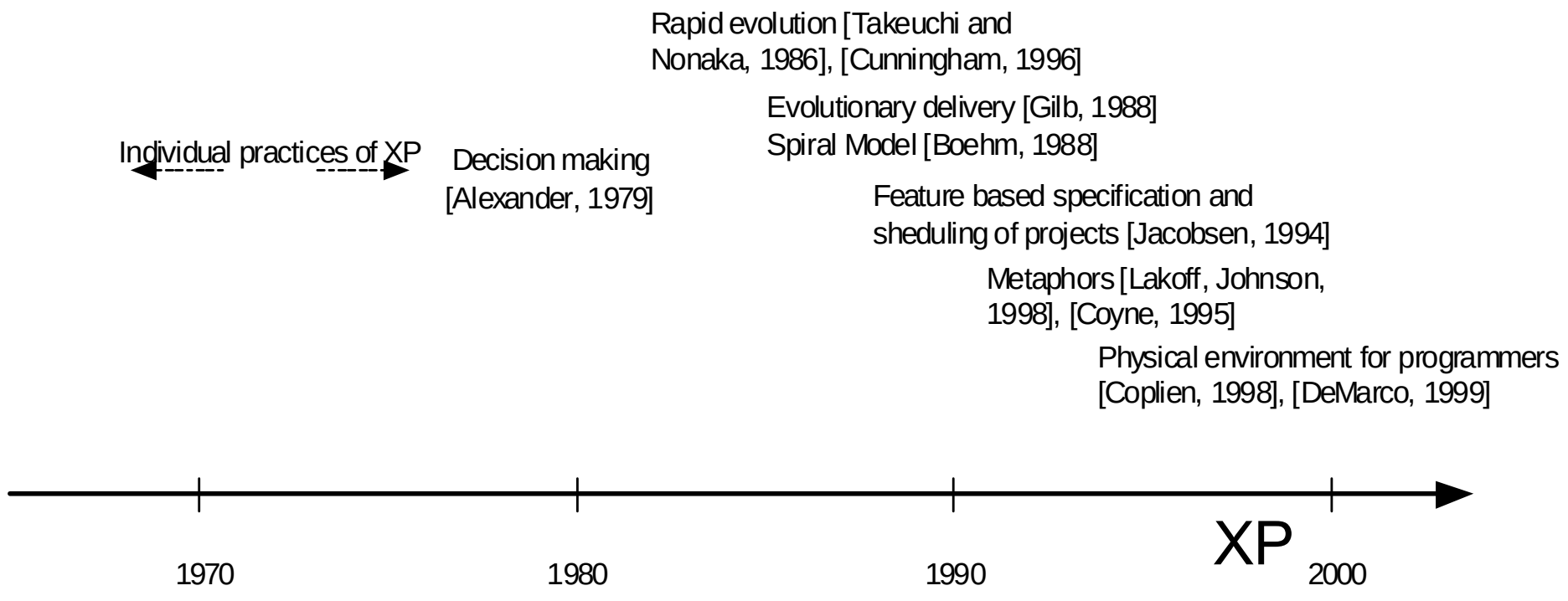

*Figure 3. Roots of Extreme Programming*

XP aims at enabling successful software development despite vague or constantly changing requirements in small to medium sized teams. Short iterations with small releases and rapid feedback, customer participation, communication and coordination, continuous integration and testing, collective ownership of the code, limited documentation and pair programming are among the main characteristics of XP. The practices of XP are presented in the following according to Beck (1999a; 2000)

**Planning game**
Close interaction between the customer and the programmers. The programmers estimate the effort needed for the implementation of customer stories and the customer then decides about the scope and timing of releases.

**Small/short releases**
A simple system is "productionized" (see Figure 2) rapidly - at least once in every 2 to 3 months. New versions are then released even daily, but at least monthly.

**Metaphor**
The system is defined by a metaphor/set of metaphors between the customer and the programmers. This "shared story" guides all development by describing how the system works.

**Simple design**
The emphasis is on designing the simplest possible solution that is implementable at the moment. Unnecessary complexity and extra code are removed immediately.

**Testing**
Software development is test driven. Unit tests are implemented before the code and are run continuously. Customers write the functional tests.

**Refactoring**
Restructuring the system by removing duplication, improving communication, simplifying and adding flexibility.

**Pair programming**
Two people write the code at one computer.

**Collective ownership**
Anyone can change any part of the code at any time.

**Continuous integration**
A new piece of code is integrated into the code-base as soon as it is ready. Thus, the system is integrated and built many times a day. All tests are run and they have to be passed for the changes in the code to be accepted.

**40-hour week**
A maximum of 40-hour working week. No two overtime weeks in a row are allowed. If this happens, it is treated as a problem to be solved.

**On-site customer**
Customer has to be present and available full-time for the team.

**Coding standards**

Coding rules exist and are followed by the programmers. Communication through the code should be emphasized.

**Open workspace**
A large room with small cubicles is preferred. Pair programmers should be placed in the center of the space.

**Just rules**
Team has its own rules that are followed, but can also be changed at any time. The changes have to be agreed upon and their impact has to be assessed.

#### 3.1.4. Adoption and experiences

Beck suggests (1999a, p. 77) that XP should be adopted gradually.

> *"If you want to try XP, for goodness sake don't try to swallow it all at once. Pick the worst problem in your current process and try solving it the XP way."*

One of the fundamental ideas of XP is that there is no process that fits every project as such, but rather practices should be tailored to suit the needs of individual projects (Beck 2000). The question is how far the individual practices and principles can be stretched so that we can still talk about practicing XP? And just how much can we leave out? In fact, there are no experience reports in which all the practices of XP have been adopted. Instead, a partial adoption of XP practices, as suggested by Beck, has been reported on several occasions (Grenning 2001; Schuh 2001)

Practical viewpoints for adopting XP have been documented in *Extreme Programming Installed* (Jeffries et al. 2001). The book describes a collection of techniques, covering most XP practices, mostly elaborated during an extensive industrial software project, where XP was used. A recurring theme in the book is the estimation of the difficulty and duration of the tasks at hand. The authors suggest using *spikes* as a means to achieve this. A spike is a short throwaway experiment in code, used to get a grasp on how a particular programming problem might be solved. The adoption process itself is not discussed in the book, but the techniques covered lower the adoption threshold by giving concrete examples of what it is like to actually perform XP.

XP is the most documented one of the different agile methods and it has triggered new research, articles and experience reports on the individual XP practices, such as pair programming (e.g.,Williams et al. 2000; Haungs 2001), as well as on applying the method itself. Mostly successful experiences (e.g., Anderson et al. 1998; Schuh 2001) of applying XP have been reported. These studies provide insight on the possibilities and restrictions of XP. Maurer and Martel (2002a) include some concrete numbers regarding the productivity gains using XP in a web development project. They report an average increase of 66.3% in the new lines of code produced, 302.1% increase in the number of new methods developed and 282.6% increase in the number of new classes implemented in a development effort. More details of their measures used and the results obtained can be found in (Maurer and Martel 2002b).

### 3.1.5. Scope of use

As stated by Beck (2000), the XP methodology is by no means suitable everywhere, nor have all its limits yet been identified. This calls for more empirical and experimental research on the subject from different perspectives. However, some limits have been identified.

XP is aimed for small and medium sized teams. Beck (2000) suggests the team size to be limited between three and a maximum of twenty project members. The physical environment is also important in XP. Communication and coordination between project members should be enabled at all times. For example, Beck (2000) mentions that a scattering of programmers on two floors or even on one floor is intolerable for XP. However, the geographical distribution of teams is not necessarily outside the scope of XP in case it includes "two teams working on related projects with limited interaction (Beck 2000, 158).

The business culture affecting the development unit is another focal issue in XP. Any resistance against XP practices and principles on behalf of project members, management or customer may be enough to fail the process (Beck 2000). Also technology might provide insuperable obstacles for the success of an XP project. For example, a technology that does not support "graceful change" or demands a long feedback time is not suitable for XP processes.

### 3.1.6. Current research

Research on XP is growing. There are many published papers on various aspects of XP, but probably due to it being seen more as a practical rather than an academic method, most papers focus on experiences of using XP in various scopes, and empirical findings on its practices. Some of these papers can be found in, e.g., (Succi and Marchesi 2000).

## 3.2. Scrum

The first references in the literature to the term 'Scrum' point to the article of Takeuchi and Nonaka (1986) in which an adaptive, quick, self-organizing product development process originating from Japan is presented (Schwaber and Beedle 2002). The term 'scrum' originally derives from a strategy in the game of rugby where it denotes "getting an out-of play ball back into the game" with teamwork (Schwaber and Beedle 2002).

The Scrum approach has been developed for managing the systems development process. It is an empirical approach applying the ideas of industrial process control theory to systems development resulting in an approach that reintroduces the ideas of flexibility, adaptability and productivity (Schwaber and Beedle 2002). It does not define any specific software development techniques for the implementation of a software. Scrum concentrates on how the team members should function in order to produce the system flexibly in a constantly changing environment.

The main idea of Scrum is that systems development involves several environmental and technical variables (e.g. requirements, time frame, resources, and technology) that are likely to change during the process. This makes the development process unpredictable and complex, requiring flexibility of the systems development process for it to be able to respond to the changes. As a result of the development process, a system is produced which is useful when delivered (Schwaber 1995).

Scrum helps to improve the existing engineering practices (e.g. testing practices) in an organization, for it involves frequent management activities aiming at consistently identifying any deficiencies or impediments in the development process as well as the practices that are used.

### 3.2.1. Process

Srcum process includes three phases: pre-game, development and post-game (Figure 4).

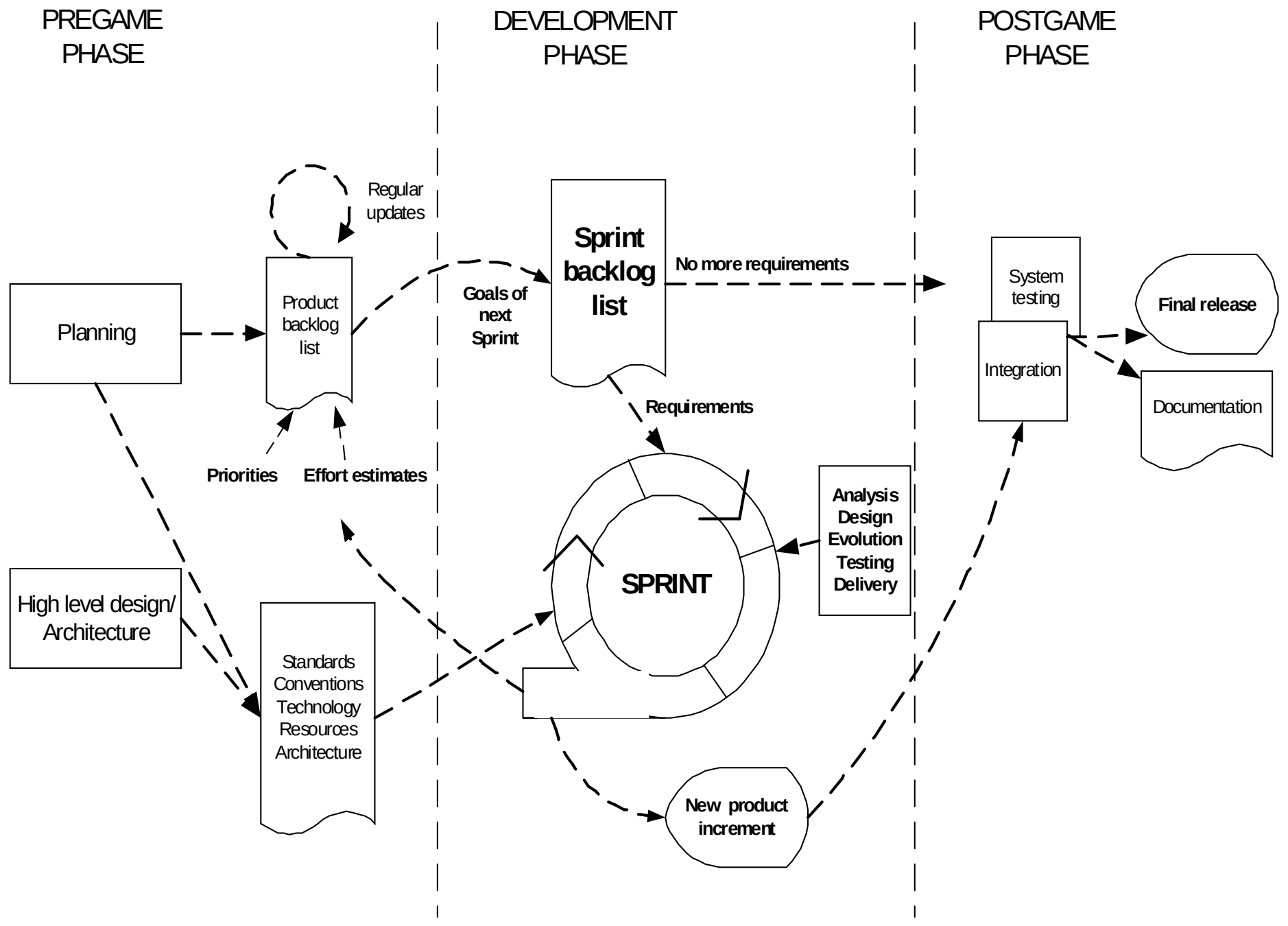

*Figure 4. Scrum Process*

In the following, the Scrum phases are introduced according to Schwaber (1995; 2002).

**The pre-game phase** includes two sub-phases: Planning and Architecture/High level design (Figure 4).

> Planning includes the definition of the system being developed. A **Product Backlog list** (see 3.2.3) is created containing all the requirements that are currently known. The requirements can originate from the customer, sales and marketing division, customer support or software developers. The requirements are prioritized and the effort needed for their implementation is estimated. The product Backlog list is constantly updated with new and more detailed items, as well as with more accurate estimations and new priority orders. Planning also includes the definition of the project team, tools and other resources, risk assessment and controlling issues, training needs and verification management approval. At every iteration, the updated product Backlog is reviewed by the Scrum Team(s) so as to gain their commitment for the next iteration.
>
> In the architecture phase, the high level design of the system including the architecture is planned based on the current items in the Product Backlog. In case of an enhancement to an existing system, the changes needed for implementing the Backlog items are identified along with the problems they may cause. A design review meeting is held to go over the proposals for the implementation and decisions are made on the basis of this review. In addition, preliminary plans for the contents of releases are prepared.

**The development phase** (also called the game phase) is the agile part of the Scrum approach. This phase is treated as a "black box" where the unpredictable is expected. The different environmental and technical variables (such as time frame, quality, requirements, resources, implementation technologies and tools, and even development methods) identified in Scrum, which may change during the process, are observed and controlled through various Scrum practices during the Sprints (see the section below) of the development phase. Rather than taking these matters into consideration only at the beginning of the software development project, Scrum aims at controlling them constantly in order to be able to flexibly adapt to the changes.

In the development phase the system is developed in **Sprints** (Figure 4 and 3.2.3). Sprints are iterative cycles where the functionality is developed or enhanced to produce new increments. Each Sprint includes the traditional phases of software development: requirements, analysis, design, evolution and delivery (Figure 4) phases. The architecture and the design of the system evolve during the Sprint development. One Sprint is planned to last from one week to one month. There may be, for example, three to eight Sprints in one systems development process before the system is ready for distribution. Also there may be more than one team building the increment.

The **post-game phase** contains the closure of the release. This phase is entered when an agreement has been made that the environmental variables such as the requirements are completed. In this case, no more items and issues can be found nor can any new ones be invented. The system is now ready for the release and the preparation for this is done during the post-game phase, including the tasks such as the integration, system testing and documentation (Figure 4).

### 3.2.2. Roles and responsibilities

There are six identifiable roles in Scrum that have different tasks and purposes during the process and its practices: Scrum Master, Product Owner, Scrum Team, Customer, User and Management. In the following, these roles are presented according to the definitions of Schwaber and Beedle (2002):

**Scrum Master**
Scrum Master is a new management role introduced by Scrum. Scrum Master is responsible for ensuring that the project is carried through according to the practices, values and rules of Scrum and that it progresses as planned. Scrum Master interacts with the project team as well as with the customer and the management during the project. He is also responsible for ensuring that any impediments are removed and changed in the process to keep the team working as productively as possible.

**Product Owner**
Product Owner is officially responsible for the project, managing, controlling and making visible the Product Backlog list. He is selected by the Scrum Master, the customer and the management. He makes the final decisions of the tasks related to product Backlog (see 3.2.3), participates in estimating the development effort for Backlog items and turns the issues in the Backlog into features to be developed.

**Scrum Team**
Scrum Team is the project team that has the authority to decide on the necessary actions and to organize itself in order to achieve the goals of each Sprint. The scrum team y is involved, for example, in effort estimation, creating the Sprint Backlog (see 3.2.3), reviewing the product Backlog list and suggesting impediments that need to be removed from the project.

**Customer**
Customer participates in the tasks related to product Backlog items (see 3.2.3) for the system being developed or enhanced.

**Management**
Management is in charge of final decision making, along with the charters, standards and conventions to be followed in the project. Management also participates in the setting of goals and requirements. For example, the management is involved in selecting the Product Owner, gauging the progress and reducing the Backlog with Scrum Master.

### 3.2.3. Practices

Scrum does not require or provide any specific software development methods/practices to be used. Instead, it requires certain management practices and tools in the various phases of Scrum to avoid the chaos caused by unpredictability and complexity (Schwaber 1995).

In the following, the description of Scrum practices is given based on Schwaber and Beedle (2002).

**Product Backlog**
Product Backlog defines everything that is needed in the final product based on current knowledge. Thus, Product Backlog defines the work to be done in the project. It comprises a prioritized and constantly updated list of business and technical requirements for the system being build or enhanced. Backlog items can include, for example, features, functions, bug fixes, defects, requested enhancements and technology upgrades. Also issues requiring solution before other Backlog items can be done are included in the list. Multiple actors can participate in generating Product Backlog items, such as customer, project team, marketing and sales, management and customer support.

This practice includes the tasks for creating the Product Backlog list, and controlling it consistently during the process by adding, removing, specifying, updating, and prioritizing Product Backlog items. The Product Owner is responsible for maintaining the Product Backlog.

**Effort estimation**

Effort estimation is an iterative process, in which the Backlog items estimates are focused on a more accurate level when more information is available on a certain Product Backlog item. The Product Owner together with the Scrum Team(s) are responsible for making the effort estimation.

**Sprint**

Sprint is the procedure of adapting to the changing environmental variables (requirements, time, resources, knowledge, technology etc.). The Scrum Team organizes itself to produce a new executable product increment in a Sprint that lasts approximately thirty calendar days. The working tools of the team are Sprint Planning Meetings, Sprint Backlog and Daily Scrum meetings (see below). The Sprint with its practices and inputs is illustrated in Figure 5.

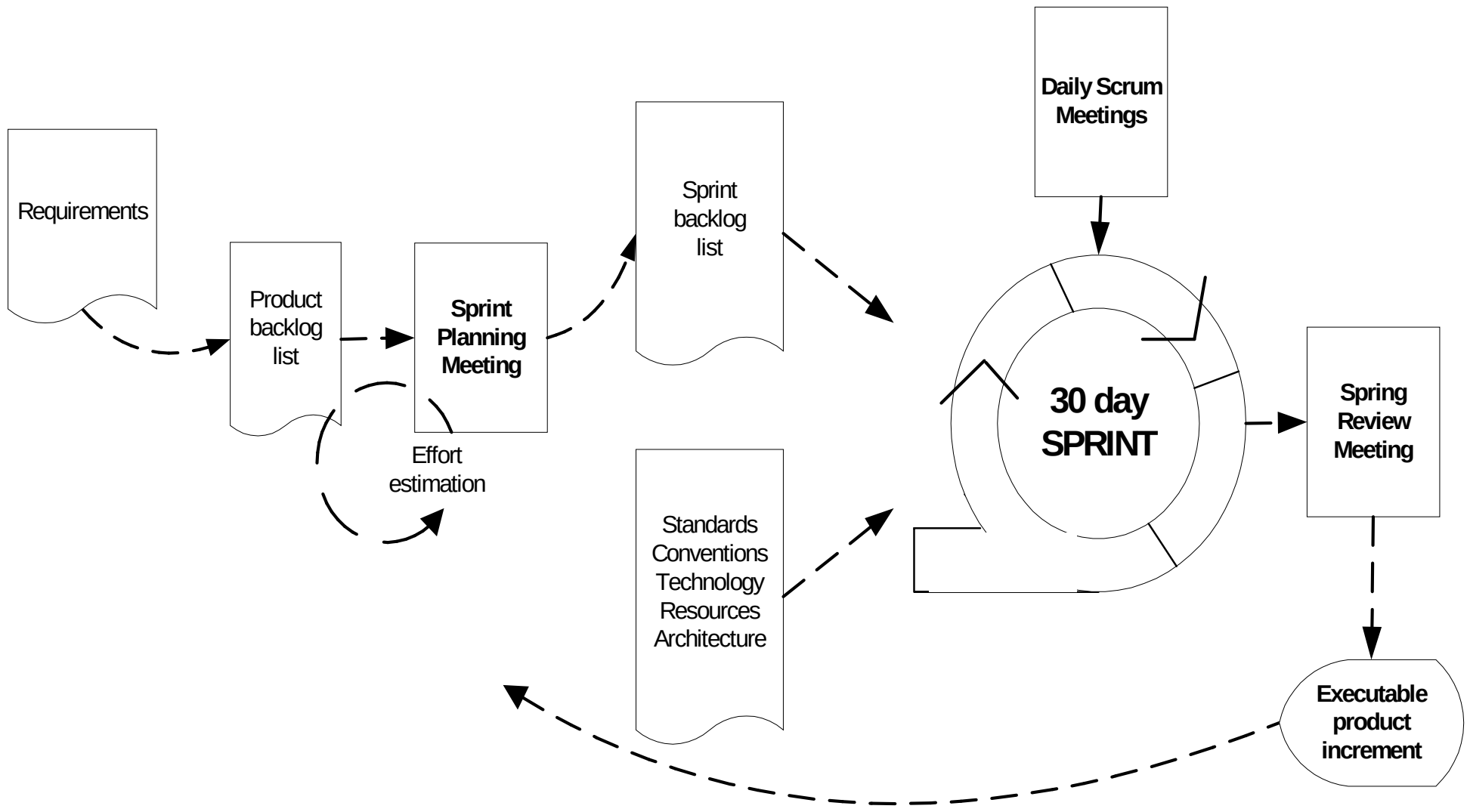

*Figure 5. Practices and Inputs of Sprint*

**Sprint Planning meeting**

A Sprint Planning Meeting is a two-phase meeting organized by the Scrum Master. The customers, users, management, Product Owner and Scrum Team participate in the first phase of the meeting to decide upon the goals and the functionality of the next Sprint (see Sprint Backlog below). The second phase of the meeting is held by the Scrum Master and the Scrum Team focusing on how the product increment is implemented during the Sprint.

**Sprint Backlog**

Sprint Backlog is the starting point for each Sprint. It is a list of Product Backlog items selected to be implemented in the next Sprint. The items are selected by the Scrum Team together with the Scrum Master and the Product Owner in the Sprint Planning meeting, on the basis of the prioritized items (see 3.2.3) and goals set for the Sprint. Unlike the Product Backlog, the Sprint Backlog is stable until the Sprint (i.e. 30 days) is completed. When all the items in the Sprint Backlog are completed, a new iteration of the system is delivered.

**Daily Scrum meeting**

Daily Scrum meetings are organized to keep track of the progress of the Scrum Team continuously and they also serve as planning meetings: what has been done since the last meeting and what is to be done before the next one. Also problems and other variable matters are discussed and controlled in this short (approximately 15 minutes) meeting held daily. Any deficiencies or impediments in the systems development process or engineering practices are looked for, identified and removed to improve the process. The Scrum Master conducts the Scrum meetings. Besides the Scrum team also the management, for example, can participate in the meeting.

**Sprint Review meeting**

On the last day of the Sprint, the Scrum Team and the Scrum Master present the results (i.e. working product increment) of the Sprint to the management, customers, users and the Product Owner in an informal meeting. The participants assess the product increment and make the decision about the following activities. The review meeting may bring out new Backlog items and even change the direction of the system being built.

### 3.2.4. Adoption and experiences

Since Scrum does not require any specific engineering practices, it can be adopted to manage whatever engineering practices are used in an organization. (Schwaber and Beedle 2002). However, Scrum may change the job descriptions and customs of the Scrum project team considerably. For example, the project manager, i.e. the Scrum Master, does no longer organize the team, but the team organizes itself and makes the decisions concerning what to do. This can be called a self organizing team (Schwaber and Beedle 2002). Instead, the Scrum Master works to remove the impediments of the process, runs and makes the decisions in the Daily Scrums and validates them with the management. His role is now more that of a coach rather than manager in the project.

Schwaber and Beedle (2002) identify two types of situations in which Scrum can be adopted: an existing project and a new project. These are described in the following. A typical case of adopting Scrum in an existing project is a situation where the development environment and the technology to be used exist, but the project team is struggling with problems related to changing requirements and complex technology. In this case, the introduction of Scrum is started with Daily Scrums with a Scrum Master. The goal of the first Sprint should be to "demonstrate any piece of user functionality on the selected technology" (Schwaber and Beedle 2002, p. 59). This will help the team to believe in itself, and the customer to believe in the team. During the Daily Scrums of the first Sprint, the impediments of the project are identified and removed to enable progress for the team. At the end of the first Sprint, the customer and the team together with the Scrum Master hold a Sprint Review and together decide where to go next. In case of carrying on with the project, a Sprint Planning meeting is held to decide upon the goals and requirements for the following Sprint.

In case of adopting Scrum in a new project, Schwaber and Beedle (2002) suggest first working with the team and the customer for several days to build an initial Product Backlog. At this point, the Product Backlog may consist of business functionality and technology requirements. The goal of the first Sprint is then to "demonstrate a key piece of user functionality on the selected technology" (Schwaber and Beedle 2002, p. 59). This, naturally, first requires designing and building an initial system framework, i.e. a working structure for the system to be built, and to which new features can be added. The Sprint

Backlog should include the tasks necessary for reaching the goal of the Sprint. As the main issue at this point concerns the adoption of Scrum, the Sprint Backlog also includes the tasks of setting up the team and Scrum roles, and building management practices in addition to the actual tasks of implementing the demo. As the team is working with the Sprint Backlog, the Product Owner works with the customer to build a more comprehensive Product Backlog so that the next Sprint can be planned after the first Sprint Review is held.

Rising and Janof (2000) report successful experiences of using Scrum in three software development projects. They suggest that "Clearly, [Scrum] is not an approach for large, complex team structures, but we found that even small, isolated teams on a large project could make use of some elements of Scrum. This is true process diversity". The interfaces between the smaller subteams must then be clearly defined. Better results were obtained by tailoring some of the Scrum principles such as the daily Scrum meeting. Teams in Rising and Janof's cases decided that two to three times per week is sufficient enough to keep the Sprint in schedule. Other positive experiences include, for example, increased volunteerism within the team and more efficient problem resolution.

### 3.2.5. Scope of use

Scrum is a method suitable for small teams of less than 10 engineers. Schwaber and Beedle suggest for the team to comprise five to nine project members (2002). If more people are available, multiple teams should be formed.

### 3.2.6. Current research

Recently, efforts to integrate XP and Scrum[8] together can be found. Scrum is seen to provide the project management framework, which is supported by the XP practices to form an integrated package for software development teams. Authors claim that this increases xp scalability to larger projects. However, no studies exist that would support their arguments.

---

[8] xp@Scrum, www.controlchaos.com/xpScrum.htm, 16.8.2002

## 3.3. Crystal family of methodologies

The Crystal family of methodologies includes a number of different methodologies for selecting the most suitable methodology for each individual project. Besides the methodologies, the Crystal approach also includes principles for tailoring the methodologies to fit the varying circumstances of different projects.

Each member of the Crystal family is marked with a color indicating the 'heaviness' of the methodology, i.e. the darker the color the heavier the methodology. Crystal suggests choosing the appropriate color of methodology for a project based on its size and criticality (Figure 6). Larger projects are likely to ask for more coordination and heavier methodologies than smaller ones. The more critical the system being developed the more rigor is needed. The character symbols in the Figure 6 indicate a potential loss caused by a system failure (i.e. the criticality level): **C**omfort (C), **D**iscretionary money (D), **E**ssential money (E) and **L**ife (L) (Cockburn 2002a). In other words, criticality level C indicates that the system crash on defects causes a loss of comfort for the user whereas defects in a life critical system may literally cause loss of life.

The dimensions of criticality and size are represented by a project category symbol described in Figure 6; thus, for example, D6 denotes a project with a maximum of six persons delivering a system of maximum criticality of discretionary money.

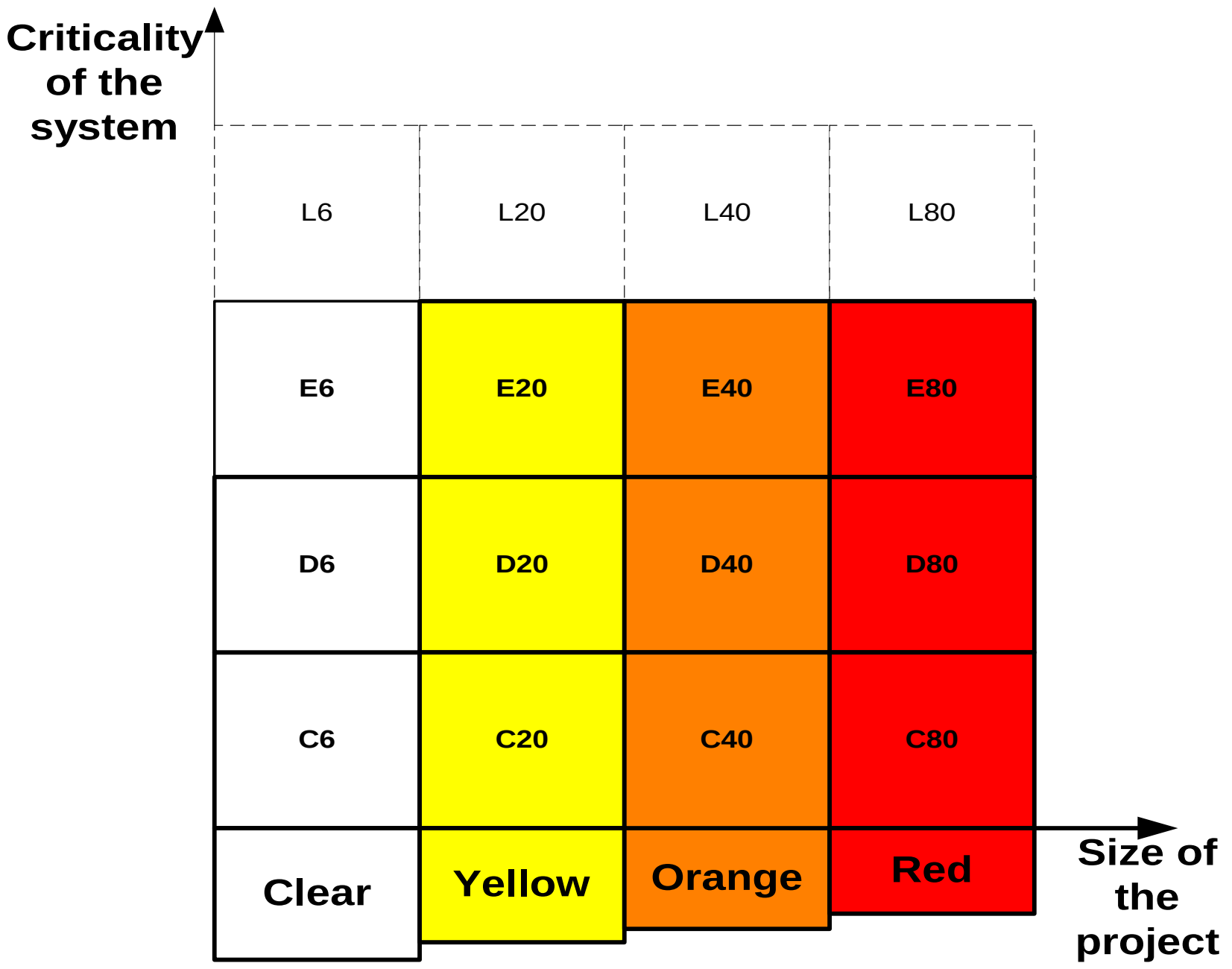

*Figure 6. Dimensions of Crystal methodologies* (Cockburn 2002a)

There are certain rules, features and values that are common to all the methods in the Crystal family. First of all, the projects always use incremental development cycles with a maximum increment length of four months, but preferably between one and three months (Cockburn 2002a). The emphasis is on communication and cooperation of people. Crystal methodologies do not limit any development practices, tools or work products, while also allowing the adoption of, for example, XP and Scrum practices (Cockburn 2002a). Furthermore, the Crystal approach embodies objectives for reducing intermediate work products and shaping conventions within a methodology for individual projects and developing them as the project evolves.

There are currently three main Crystal methodologies constructed: Crystal Clear, Crystal Orange and Crystal Orange Web (Cockburn 2002a). The first two of these have been experimented in practice and thus also introduced and discussed in this section.

### 3.3.1. Process

All of the methodologies of the Crystal family provide guidelines of policy standards, work products, "local matters", tools, standards and roles (see 3.3.2) to be followed in the development process. Crystal Clear and Crystal Orange are the two Crystal family members that have been constructed and used (Cockburn 1998; Cockburn 2002a). Crystal Orange (Cockburn 1998) also presents the activities included in the process. In this subsection, these two methodologies are discussed by viewing their contexts, similarities and differences and illustrating the process and activities of Crystal Orange.

Crystal Clear is designed for very small projects (D6 project category projects), comprising up to six developers. However, with some extension to communication and testing it can be applied also to E8/D10 projects. A team using Crystal Clear should be located in a shared office-space due to the limitations in its communication structure (Cockburn 2002a).

Crystal Orange is designed for medium-sized projects, with a total of 10 to 40 project members (D40 category), and with a project duration of one to two years. Also E50 category projects may be applicable within Crystal Orange with additions to its verification-testing processes. (Cockburn 2002a). In Crystal Orange, the project is split up for several teams with cross-functional groups (see 3.3.2) using the Holistic Diversity strategy (see 3.3.3). Still, this method does not support the distributed development environment. The Crystal Orange emphasizes the importance of the time-to-market issue. The trade-offs between extensive deliverables and rapid change in requirements and design results in a limited number of deliverables enabling to reduce the cost of maintaining them, but still keeping the communication functioning efficiently between the teams (Cockburn 1998).

In the following, the differences and similarities between the Crystal Clear and Orange methods are discussed regarding the different elements provided by the Crystal family for each of its members. The roles are further discussed in sub-section 3.3.2.

### Policy standards

Policy standards are the practices that need to be applied during the development process. Both the Crystal Clear as well as Crystal Orange suggest the following policy standards (Cockburn 2002a):

- Incremental delivery in regular basis
- Progress tracking by milestones based on software deliveries and major decisions rather than written documents
- Direct user involvement
- Automated regression testing of functionality
- Two user viewing per release
- Workshops for product- and methodology-tuning at the beginning and in the middle of each increment

The only difference between the policy standards of these two methodologies is that Crystal Clear suggest incremental delivery within a two to three month time frame whereas in Crystal Orange the increments can be extended to four months at the maximum.

The policy standards of these Crystal methodologies are mandatory but can, however, be replaced with equivalent practices of other methodologies such as XP or Scrum (Cockburn 2002a).

### Work products

Cockburn's (2002a) requirements for work products of Crystal Clear and Crystal Orange differ to a certain extent. The similar work products of both Crystal Clear and Orange include the following items: release sequence, common object models, user manual, test cases, and migration code.

In addition, Crystal Clear includes annotated use cases/feature descriptions, whereas in Crystal Orange the requirements document is required. In Crystal

Clear, the schedule is documented only on user viewings and deliveries and the comparable work product in Orange that Cockburn lists is (2002a) "schedule", indicating a need for more extensive scheduling of the project. The more lightweight nature of the work products in Crystal Clear emerges also in design documentation: while Orange demands UI design documents and inter-team specifications, in Clear only screen drafts and design sketches and notes are suggested if needed. In addition to these work products, Crystal Orange requires status reports.

### Local matters

Local matters are the procedures of Crystal that have to be applied, but their realization is left up to the project itself. The local matters of Crystal Clear and Crystal Orange do not largely differ from each other. Both methodologies suggest that templates for the work products as well as coding, regression testing and user interface standards should be set and maintained by the team itself (Cockburn 2002a). For example, project documentation is required but their content and form remains a local matter. Above this, also the techniques to be used by the individual roles in the project are not defined by Crystal Clear nor Crystal Orange.

### Tools

The tools that the Crystal Clear methodology requires are compiler, versioning and configuration-management tool, and printing whiteboards (Cockburn 2002a). Printing whiteboards are used, for example, for replacing the formal design documents and meeting summaries. In other words, they are used for storing and presenting material that otherwise should be typed down in a formal documentation format after a meeting.

The minimum tools required by Crystal Orange are those used for versioning, programming, testing, communication, project tracking, drawing, and performance measuring. In addition, screen drivers are needed for repeatable GUI tests. (Cockburn 1998).

### Standards

Crystal Orange proposes selecting the notation standards, design conventions, formatting standards and quality standards (Cockburn 1998) to be used in the project.

### Activities

The activities of Crystal Orange are discussed in more detail in section 3.3.3. However, they are included in the Figure 7 as a part of the illustration of one increment of Crystal process.

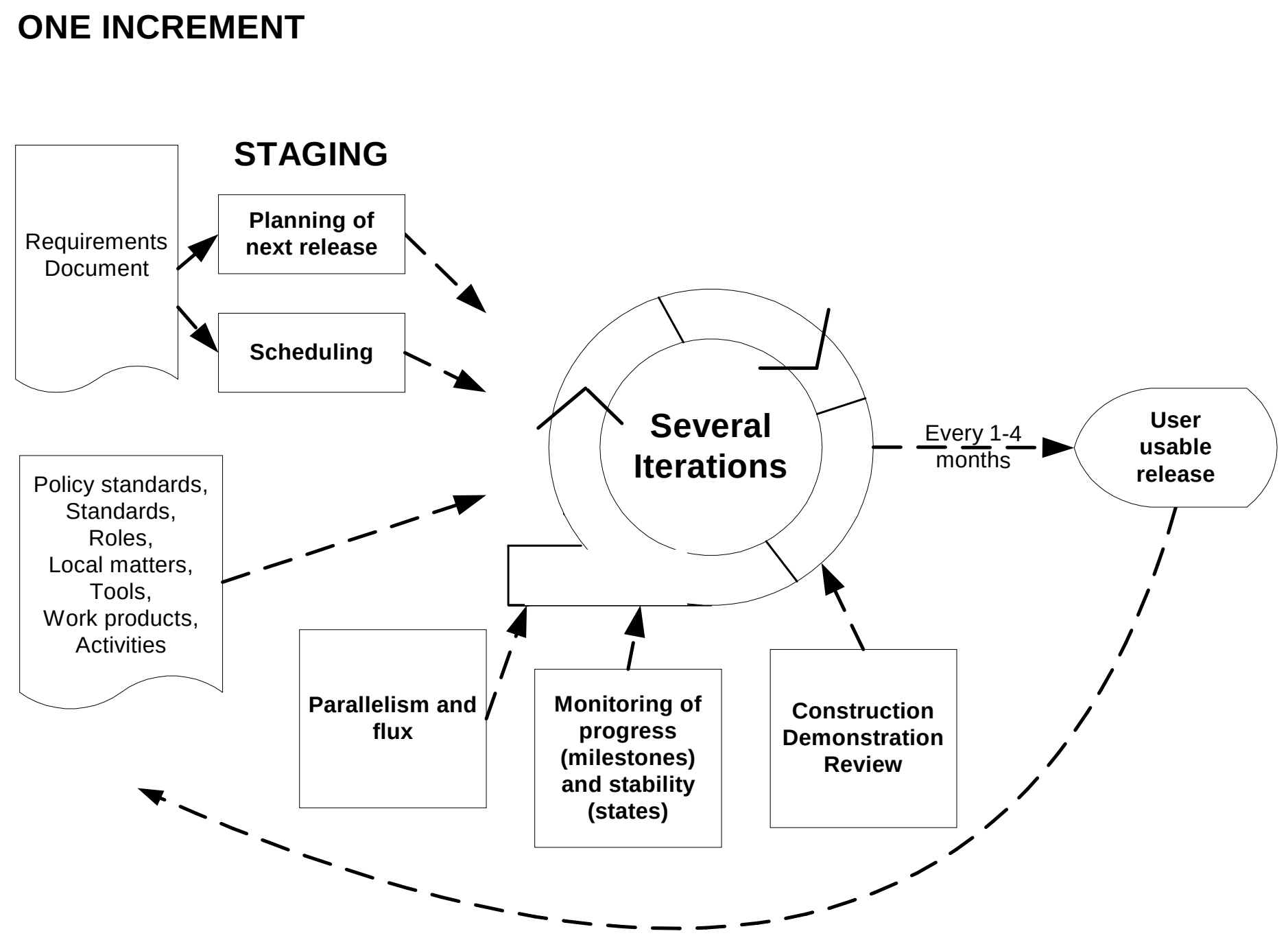

*Figure 7. One Crystal Orange Increment*

### 3.3.2. Roles and responsibilities

In this section, the roles of Crystal Clear and Crystal Orange are presented according to Cockburn (2002a). The basic difference between Crystal Clear and Orange is that in the former there is only one team in a project. In Crystal Orange there are multiple teams to follow though the project. In both methodologies, one job assignment may include multiple roles.

In Crystal Clear the main roles requiring separate persons are (Cockburn 2002a): sponsor, senior designer-programmer, designer-programmer and user. These roles embody multiple sub-roles. For example, the designer-programmer consists of business class designer, programmer, software documenter and unit tester (Cockburn 1998). The sub-roles that can be assigned to other roles are coordinator, business expert and requirements gatherer (Cockburn 2002a). The business expert represents a business view in the project, possessing knowledge about the specific business context. He should be able to take care of the business plan, paying attention to what is stable and what is changing (Cockburn 1998).

In addition to the roles introduced in Crystal Clear, Crystal Orange suggest a wide range of main roles needed in the project. The roles are grouped into several teams, such as system planning, project mentoring, architecture, technology, functions, infrastructure and external test teams (Cockburn 2002a). The teams are further divided into cross-functional groups containing different roles. (Cockburn 2002a).

Crystal Orange introduces several new roles (Cockburn 1998; Cockburn 2002a), such as UI designer, database designer, usage expert, technical facilitator, business analyst/designer, architect, design mentor, reuse point, writer, and tester. Crystal Orange (Cockburn 1998) also involves the skills and techniques needed in order to succeed in these roles. One project member can hold multiple roles. For example, reuse point is a part-time role that can be carried out by an architect or a designer-programmer, for instance (Cockburn 1998). This role refers to the identification of reusable software components and acquisition of commercial components. Other roles that need further clarification are the writer and the business analyst-designer. The writer's role includes the construction of external documentation, such as screen specifications and user manuals (Cockburn 1998). The Business analyst-designer communicates and negotiates with the users in order to specify the requirements and interfaces, and to review the design.

Each group consists of at least a business analyst designer, a UI designer, two to three designer-programmers, a database designer and possibly also a tester. Also technical facilitators may be needed in a group. The reason for having cross-functional groups is to reduce deliverables and to enhance local communication. (Cockburn 2002a)

### 3.3.3. Practices

All Crystal methodologies involve a number of practices, such as incremental development. In the description of Crystal Orange (Cockburn 1998), the increment includes activities such as staging, monitoring, reviewing, along with parallelism and flux (Figure 7). Also other activities and practices can be identified. These are included in the following descriptions.

**Staging**
Staging includes the planning of the next increment of the system. It should be scheduled to produce a working release in every three or four months (Cockburn 1998) at the maximum. A schedule of one to three months is preferred (Cockburn 2002a). The team selects the requirements to be implemented in the increment and schedules what they feel they are able to deliver (Cockburn 1998).

**Revision and review**
Each increment includes several iterations. Each iteration includes the following activities: construction, demonstration and review of the objectives of the increment (Cockburn 1998).

**Monitoring**
The progress is monitored regarding the team deliverables during the development process with respect to their progress and stability (Cockburn 1998). The progress is measured by milestones (start, review 1, review 2, test, deliver) and stability stages (wildly fluctuating, fluctuating and stable enough to review). Monitoring is required in both Crystal Clear and Crystal Orange.

**Parallelism and flux**

Once the monitoring of stability gives the result "stable enough to review" for the deliverables the next task can start. In Crystal Orange, this means that the multiple teams can proceed with maximum parallelism successfully. To ensure this, the monitoring and architecture teams review their work plans, stability and synchronization. (Cockburn 1998).

**Holistic diversity strategy**

Crystal Orange includes a method called holistic diversity strategy for splitting large functional teams into cross-functional groups. The central idea of this is to include multiple specialties in a single team (Cockburn 1998, p.214). The holistic diversity strategy also allows forming small teams with the necessary special know-how, and also considers issues such as locating the teams, communication and documentation and coordination of multiple teams. (Cockburn 1998).

**Methodology-tuning technique**

The methodology-tuning technique is one of the basic techniques of Crystal Clear and Orange. It uses project interviews and team workshops for working out a specific Crystal methodology for each individual project (Cockburn 2002a). One of the central ideas of incremental development is to allow fixing or improving the development process (Cockburn 1998). In every increment, the project can learn and use the knowledge it has gained for developing the process for the next increment.

**User viewings**

Two user viewings are suggested for Crystal Clear per one release (Cockburn 2002a). In Crystal Orange, users reviews should be organized three times for each increment (Cockburn 1998).

**Reflection workshops**

Both Crystal Clear and Orange include a rule that a team should hold pre- and post-increment reflection workshops (with recommendation for also mid-increment reflection workshops) (Cockburn 2002a).

Crystal Clear and Crystal Orange do not define any specific practices or techniques to be used by the project members in their software development tasks. The adoption of practices from other methodologies such as XP and Scrum is allowed in Crystal to replace some of its own practices, such as reflection workshops, for instance.

### 3.3.4. Adoption and experiences

At least one experience report (Cockburn 1998) can be found of a project adopting, evolving and using practices that currently form the Crystal Orange methodology (Cockburn 2002b). The two-year 'project Winifred' was a medium sized project with 20 to 40 staff members, with an incremental development strategy, an object-oriented approach and a goal of delivering a rewritten legacy mainframe system.

According to Cockburn (1998) the project Winifred run into problems in the first increment of the project. Problems with the communication within and across teams, a long requirements phase that prevented designing any architecture for several months, high turnover of technical leaders and unclear job assignments gave rise to the evolvement of software development process into a form that is now called Crystal Orange. One solution was the adoption of an iterative process for each increment. This provided the teams with a chance to identify their own weaknesses and to reorganize themselves. Also, the iterations included functional viewings with the users for defining the final requirements. This kept the users consistently involved in the project. Furthermore, the lessons learned on the first increment encouraged those involved to focus on issues such as clear job assignments, planning the lines of communication, defined ownership of deliverables and management support.

It is to be noted that in the project Winifred different teams could have different numbers of iterations depending on the their tasks. The communication across the teams, for example in the form of deliverables, was then planned to suit the iterations and vice versa.

In conclusion, the key success factors of the Winifred project were the increments enabling the fixing of the process, certain key individuals, and the organization of the team into a habit of delivering (Cockburn 1998).

### 3.3.5. Scope of use

At present, the Crystal family of methodologies does not cover life-critical projects Figure 6 (Cockburn 2002a). Another restriction identified by Cockburn (2002a) is that only co-located teams can be addressed by these methodologies.

Cockburn (2002a) has also identified limitations concerning the individual methodologies used in the Crystal family. For example, Crystal Clear has a relatively restricted communication structure and is thus suitable for only for a single team located in a single office space. Furthermore, Crystal Clear lacks system validation elements and is thus not suitable for life-critical systems. Crystal Orange also requires its teams to be located in a single office building and is capable of delivering only a system maximum of discretionary money in the level of criticality. It also lacks in its sub-team structures and, as such, is suitable for projects involving up to 40 persons. It is also not very competent regarding design- and code-verification activities and thus not suitable for life-critical systems.

### 3.3.6. Current research

While only two of the four proposed Crystal methodologies exist, Alistair Cockburn[9] continues the development of his Crystal family of methodologies to cover different types of projects and problem domains. Beyond this, no research can be identified.

---

[9] crystalmethodologies.org

## 3.4. Feature Driven Development

Feature Driven Development (FDD) is an agile and adaptive approach for developing systems. The FDD approach does not cover the entire software development process, but rather focuses on the design and building phases (Palmer and Felsing 2002). However, it has been designed to work with the other activities of a software development project (Palmer and Felsing 2002) and does not require any specific process model to be used. The FDD approach embodies iterative development with the best practices found to be effective in industry. It emphases quality aspects throughout the process and includes frequent and tangible deliveries, along with accurate monitoring of the progress of the project.

FDD consists of five sequential processes and provides the methods, techniques and guidelines needed by the project stakeholders to deliver the system. Furthermore, FDD includes the roles, artifacts, goals, and timelines needed in a project. (Palmer and Felsing 2002). Unlike some other agile methodologies, FDD claims to be suitable for the development of critical systems (Palmer and Felsing 2002).

FDD was first reported in (Coad et al. 2000). It was then further developed on the basis of a work done for a large software development project by Jeff Luca, Peter Coad and Stephen Palmer.

### 3.4.1. Process

As mentioned earlier, FDD consists of five sequential processes during which the designing and building of the system is carried out (Figure 8). The iterative part of the FDD processes (Design and Build) supports agile development with quick adaptations to late changes in requirements and business needs. Typically, an iteration of a feature involves a one to three week period of work for the team.

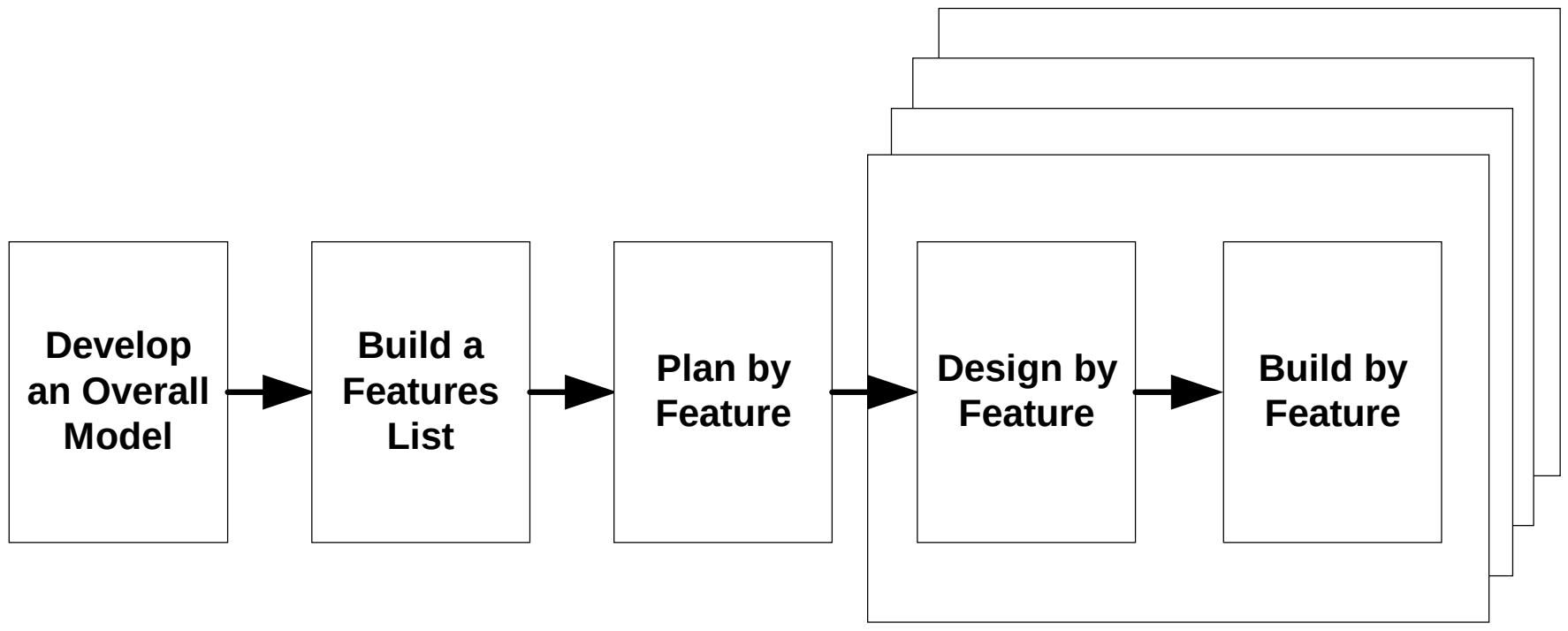

*Figure 8. Processes of FDD (Palmer and Felsing 2002)*

In the following, each of the five processes is described according to Palmer (2002).

### Develop an Overall Model

When the development of an overall model begins, the domain experts (see 3.4.2) are already aware of the scope, context and requirements of the system to be build. (Palmer and Felsing 2002). Documented requirements such as use cases or functional specifications are likely to exist at this stage. However, FDD does not explicitly address the issue of gathering and managing the requirements. The domain experts present a so called "walkthrough" in which the team members and the chief architect are informed of the high-level description of the system.

The overall domain is further divided into different domain areas and a more detailed walkthrough is held for each of them by the domain members. After each walkthrough, a development team works in small groups in order to produce object models for the domain area at hand. The development team then discusses and decides upon the appropriate object models for each of the domain areas. Simultaneously, an overall model shape is constructed for the system. (Palmer and Felsing 2002).

### Build a Features List

The walkthroughs, object models and existing requirement documentation give a good basis for building a comprehensive features list for the system being developed. In the list, the development team presents each of the **client valued functions** included in the system. The functions are presented for each of the domain areas and these function groups consist of so-called major feature sets. In addition, the major feature sets are further divided into feature sets. These represent different activities within specific domain areas. The feature list is reviewed by the users and sponsors of the system for their validity and completeness.

**Plan by Feature**

Planning by feature includes the creation of a high-level plan, in which the feature sets are sequenced according to their priority and dependencies and assigned to Chief Programmers (see 3.4.2). Furthermore, the classes identified in the "developing of an overall model" process are assigned to individual developers, i.e. class owners (see 3.4.2). Also schedule and major milestones may be set for the feature sets.

**Design by Feature and Build by Feature**

A small group of features is selected from the feature set(s) and feature teams needed for developing the selected features are formed by the class owners. The design by feature and build by feature processes are iterative procedures, during which the selected features are produced. One iteration should take from a few days to a maximum of two weeks. There can be multiple feature teams (see 3.4.2) concurrently designing and building their own set of features. This iterative process includes such tasks as design inspection, coding, unit testing, integration and code inspection. After a successful iteration, the completed features are promoted to the main build while the iteration of designing and building begins with a new group features taken from the feature set (see Figure 9).

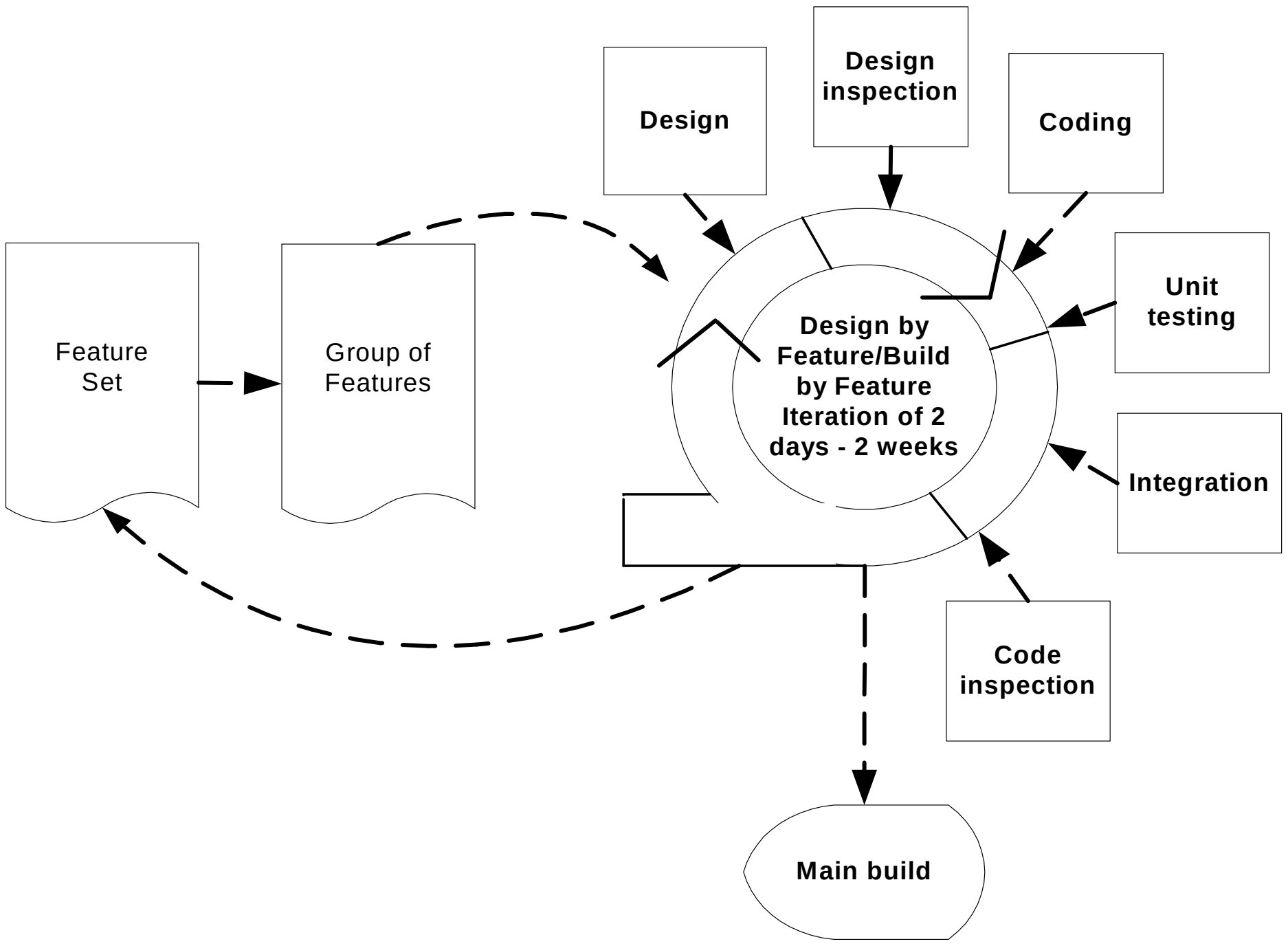

*Figure 9. The Design by feature and Build by feature processes of FDD*

### 3.4.2. Roles and responsibilities

The FDD classifies its roles into three categories: key roles, supporting roles and additional roles (Palmer and Felsing 2002). The six key roles in FDD project are: project manager, chief architect, development manager, chief programmer, class owner and domain experts. The five supporting roles comprise release manager, language lawyer/language guru, build engineer, toolsmith and system administrator. The three further roles that are needed in any project are those of the testers, deployers and technical writers. One team member can play multiple roles, and or a single role can be shared by several people (Palmer and Felsing 2002).

In the following, the tasks and responsibilities of the different roles, as presented by Palmer (2002), are described in brief.

**Project Manager**

Project manager is the administrative and financial leader of the project. One of his tasks is to protect the project team from outside distractions and to enable the team to work along with providing it with appropriate working conditions. In FDD, the project manager has the ultimate say on the scope, schedule, and staffing of the project.

**Chief Architect**

The chief designer is responsible for the overall design of the system and running the workshop design sessions held with the team (Palmer and Felsing 2002). The chief architect also makes the final decisions on all design issues. If necessary, this role can be divided into the roles of domain architect and technical architect.

**Development Manager**

The development manager leads daily development activities and solves any conflicts that may occur within the team. In addition, this role includes the responsibility of solving resourcing problems. The tasks of this role can be combined with those of the chief architect or project manager.

**Chief Programmer**

The chief programmer is an experienced developer, who participates in the requirements analysis and design of the projects. The chief programmer is responsible for leading small teams in the analysis, design and development of new features. Selecting features from the feature set(s) to be developed in the next iteration of the "design by feature and build by feature" processes and identifying the classes and class owners that are needed in the feature team for the iteration also belong to the responsibilities of the chief programmer, along with cooperation with other chief programmers in solving technical and resourcing issues, and reporting the progress of the team weekly.

**Class Owner**

Class owners work under the guidance of the chief programmer in the tasks of designing, coding, testing and documenting. He is responsible for the development of the class he has been assigned to be the owner for. Class owners form **feature teams**. For each iteration the class owners are involved whose class is included in the features selected for the next development iteration.

**Domain Expert**
The domain expert may be user, client, sponsor, business analyst or a mixture of these. His/her task is to possess the knowledge of how the different requirements for the system under development should perform. Domain experts pass this knowledge to the developers in order to ensure that the developers deliver a competent system.

**Domain Manager**
Domain manager leads the domain experts and resolves their differences of opinion concerning the requirements for the system.

**Release Manager**
Release manager controls the progress of the process by reviewing the progress reports of chief programmers and holds short progress meetings with them. He reports the progress to the project manager.

**Language Lawyer/Language Guru**
A team member responsible for possessing a thorough knowledge of, for example, a specific programming language or technology. This role is particularly important when the project team is dealing with some new technology.

**Build Engineer**
A person responsible for setting up, maintaining and running the build process, including the tasks of managing the version control system and publishing documentation.

**Toolsmith**
Toolsmith is a role for building small tools for development, test and data conversion teams in the project. Also, toolsmith may be working with setting up and maintaining databases and Web sites for project-specific purposes.

**System Administrator**
The task of a system administrator is to configure, to manage and to troubleshoot the servers, network of workstations and development and testing environments used by the project team. Also, the system administrator may be involved in the productionizing of the system being developed.

**Tester**
Testers verify that the system being produced will meet the requirements of the customer. May be an independent team of part of the project team.

**Deployer**
Deployers' work is concerned with converting existing data to the format required by the new system and participating in the deployment of new releases. May be an independent team of part of the project team.

**Technical Writer**
The user documentation is prepared by technical writers, who may form an independent team or be part of the project team.

### 3.4.3. Practices

FDD consists of set of "best practices" and the developers of the method claim that even though the selected practices are not new, the specific blend of these ingredients make the five FDD processes unique for each case. Palmer and Felsing also advocate that all the practices available should be used to get the most benefit of the method as no single practice dominates the whole process (2002). FDD involves the following practices:

- Domain Object Modeling: Exploration and explanation of the domain of the problem. Results in a framework where the features are added.

- Developing by Feature: Developing and tracking the progress through a list of small functionally decomposed and client-valued functions.

- Individual Class (Code) Ownership: Each class has a single person nominated to be the one responsible for the consistency, performance and conceptual integrity of the class.

- Feature Teams: Refers to small, dynamically formed teams

- Inspection: Refers to the use of the best-known defect-detection mechanisms.

- Regular Builds: Refers to ensuring that there is always a running, demonstrable system available. Regular builds form the baseline to which new features are added.

- Configuration Management: Enables the identification and historical tracking of the latest versions of each completed source code file.

- Progress reporting: Progress is reported based on complete work to all necessary organizational levels.

The project team must put all the above practices into use in order to comply with the FDD development rules. However, the team is allowed to adapt them according to their experience level.

### 3.4.4. Adoption and experiences

FDD was used for the first time in a development work of a large and complex banking application project in the late 90’s. According to Palmer and Felsing (Palmer and Felsing 2002), FDD is suitable for new projects starting out, projects enhancing and upgrading existing code, and projects tasked with the creation of a second version of an existing application. The authors also suggest that organizations should adopt the method gradually in small pieces and finally in its entirety as the development project proceeds.

### 3.4.5. Scope of use

The authors further state that FDD is “worthy of serious consideration by any software development organization that needs to deliver quality, business-critical software systems on time” (Palmer and Felsing 2002, p. xxiii). Unfortunately, reliable experience reports are still hard to find, even though several software consulting firms seem to advocate the method as can be easily seen on the Internet.

### 3.4.6. Current research

As there are no reliable success stories or, on the other hand, failure reports on using FDD, any research, discussions and comparisons of this sort would be of utmost interest for academics and practitioners alike, enabling them to better implement FDD in each specific case, to decide when to use FDD and not some other agile method, and when not to use FDD approach. However, this research still remains to be done. As the method is relative new, it is still evolving, and more and more supporting tools will be available in the future.

## 3.5. The Rational Unified Process

The Rational Unified Process – or RUP for short – was developed by Philippe Kruchten, Ivar Jacobsen and others at Rational Corporation to complement UML (Unified Modelling Language), an industry-standard software modeling method.

RUP is an iterative approach for object-oriented systems, and it strongly embraces use cases for modeling requirements and building the foundation for a system. RUP is inclined towards object-oriented development. It does not implicitly rule out other methods, although the proposed modeling method, UML, is particularly suited for OO development (Jacobsen et al. 1994).

### 3.5.1. Process

The lifespan of a RUP project is divided into four phases named Inception, Elaboration, Construction and Transition (see Figure 10). These phases are split into iterations, each having the purpose of producing a demonstrable piece of software. The duration of an iteration may vary from two weeks or less up to six months.

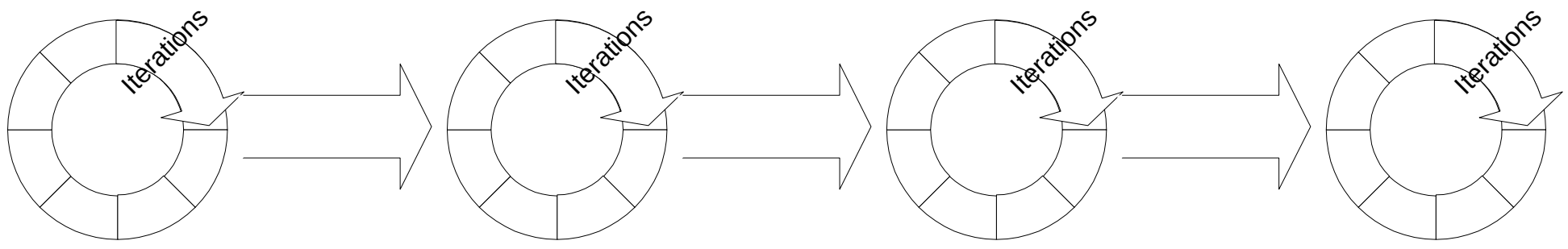

*Figure 10. RUP phases*

In the following, the RUP phases are introduced, as described in (Kruchten 2000):

In the **inception phase**, the life-cycle objectives of the project are stated so that the needs of every stakeholder (e.g. end user, purchaser, or contractor) are considered. This entails establishing, e.g., the scope and boundary conditions, and acceptance criteria of the project. Critical use cases are identified – i.e. those that will drive the functionality of the system. Candidate architectures for the system are devised, and the schedule and cost are estimated for the entire project. In addition, estimates are made for the following elaboration phase.

The **elaboration phase** is where the foundation of software architecture is laid. The problem domain is analyzed, bearing in mind the entire system that is being built. The project plan is defined in this phase – if the decision is made to proceed to the next phases at all. In order to be able to make that decision, RUP assumes that the elaboration phase will yield a sufficiently solid architecture along with sufficiently stable requirements and plans. The process, infrastructure and development environment are described in detail. As RUP emphasizes tool automation, the support for it is also established in the elaboration phase. After the phase, most use cases and all actors have been identified and described, the software architecture described, and an executable prototype of the architecture created. At the end of the elaboration phase, an analysis is made to determine the realization of risks, the stability of the vision of what the product is to become, the stability of the architecture, and the expenditure of resources versus what was initially planned.

In the **construction phase**, all remaining components and application features are developed and integrated into the product, and tested. RUP considers the construction phase a manufacturing process, where emphasis is put on managing resources and controlling costs, schedules, and quality. The results of the construction phase (alpha, beta, and other test releases) are created as rapidly as possible, while still achieving adequate quality. One or more releases are done during the construction phase, before proceeding to the final, transition phase.

The **transition phase** is entered, when the software product is mature enough (determined, for example, by monitoring the rate of system change requests) to be released to the user community. Based on user response, subsequent releases are made to correct any outstanding problems, or to finish any postponed features. The transition phase consists of beta testing, piloting, training the users and maintainers of the system, and rolling the product out to marketing, distribution, and sales teams. Several iterations are often made, with beta and general availability releases. User documentation (manuals, course material) is also produced.

Throughout the phases, nine **workflows** (Kruchten 2000), are taking place in parallel. Each iteration, more or less extensively, addresses all the nine workflows. The workflows are **Business Modelling, Requirements, Analysis & Design, Implementation, Test, Configuration & Change Management, Project Management,** and **Environment.** Most of these are self-evident. Two of them, however, are less commonly seen in other software process descriptions, which is why they are described in more detail in the following.

**Business Modelling** is used for ensuring that the customer's business needs are catered for. By analyzing the customer's organization and business processes, a better understanding of the structures and dynamics of the business is gained. The business model is built as a business use-case model, consisting of business use cases and actors. The importance of business modeling depends entirely on the purpose for which the software is built, and the phase may be entirely omitted if not deemed necessary. Business modeling is most often done during the inception and elaboration phases.

The **Environment** workflow is solely designed to support development work. The activities in this workflow include implementing and configuring the RUP itself, selecting and acquiring required tools, developing in-house tools, providing software and hardware maintenance, and training. Practically all the work in the environment workflow is done during the inception phase.

### 3.5.2. Roles and responsibilities

The way roles are assigned in RUP is activity driven, so that there is a role for each activity. RUP defines thirty roles, called workers. The casting is fairly conventional (e.g. Architect, Designer, Design Reviewer, Configuration Manager), with the exception of the roles defined in the Business modeling and Environment workflows.

In continuous co-operation with members of the business organization, a **Business-Process Analyst** leads and coordinates the defining of the business use cases, which describe the important processes and actors (e.g., customers) in the organization business. He/she thus forms a high-level abstraction of the business that is being modeled, in the form of a business use-case model, and also defines the business object model, which describes how the processes and actors interact, and creates a glossary listing the important terms used in the business.

The business use-case model is split into parts, each of which is elaborated into a business use case by a **Business Designer.** While doing so, he identifies and documents the roles and entities within the organization.

A **Business-Model Reviewer** reviews all of the artifacts produced by the Business-Process Analyst and the Business Designer.

The Environment workflow has two interesting roles: a **Course Developer** produces course material (slides, tutorials, examples, etc.) for the end users of the system that is under development. A **Toolsmith** develops in-house tools to support development, to enhance automation for tedious, repeating tasks, and to improve the integration between tools.

The number of people required in a RUP project varies considerably, depending on the scope in which the workflows are implemented. For example, the business modeling workflow (along with the corresponding roles) can be omitted entirely, if the business model has no significance in the end product.

### 3.5.3. Practices

The cornerstones of RUP can be listed in six practices that lie beneath the phase-workflow structure and the roles of RUP. The practices are listed below.

*Table 3. RUP practices (Kruchten 2000).*

| **RUP practice** | **Description** |
|---|---|
| Develop software iteratively | Software is developed in small increments and short iterations, which allows identifying risks and problems early on, and reacting to them adequately. |
| Manage requirements | Identifying the requirements of a software system change over time, and those requirements that have the greatest impact on the project goals is a continuous process. A disciplined approach for managing requirements is required, where the requirements can be prioritized, filtered and traced. Inconsistencies can then be more easily spotted. Communication has better chances of succeeding when based on clearly defined requirements. |
| Use component-based architectures | Software architecture can be made more flexible by the use of components - those parts of the system that are most likely to change can be isolated and more easily managed. In addition, building reusable components can potentially save a substantial amount of future development effort. |
| Visually model software | Models are built, because complex systems are impossible to understand in their entirety. By using a common visualization method (such as UML) among the development team, system architecture and design can be captured unambiguously, and communicated to all parties concerned. |

| RUP practice | Description |
|---|---|
| Verify software quality | By testing during every iteration, defects can be identified earlier on in the development cycle, greatly reducing the cost of fixing them. |
| Control changes to software | Any changes to requirements must be managed, and their effect on software must be traceable. The maturity of software can also be effectively measured by the frequency and types of the changes made. |

### 3.5.4. Adoption and experiences

(Kruchten 2000) claims that RUP can often be adopted, in whole or in part, “out of the box”. In many cases, however, a thorough configuration, i.e. modification of RUP, is suggested before implementing it. The configuration yields a development case, which lists all deviations that have to be made with respect to a complete RUP.

The adoption process itself is an iterative six-step program, which is repeated, until the new process has been completely implemented – which is, in essence, a plan-do-check-act cycle. Each increment brings in new practices to implement, and adjusts the previously added practices. Based on feedback from the previous cycle, the development case is updated if needed. Piloting the process first in a suitable project is suggested.

Despite the need for extensive tailoring, however, RUP has been implemented in many organizations. The success of RUP may also be, to some extent, due to the fact that Rational Software sells popular tools to support the RUP phases, for example, Rational Rose, a UML modeling tool, and ClearCase, a software configuration management tool.

### 3.5.5. Scope of use

RUP is not generally considered particularly "agile". The method was originally developed to cater for the entire software production process, and therefore contains extensive guidelines for process phases that are next to irrelevant in an environment that is likely to call for an agile approach. Recent studies (Martin 1998; Smith 2001) have, however, examined the potential of RUP for using the method in an agile manner. It appears that the adoption phase is the key to adjusting RUP towards agility. The complete RUP lists over a hundred artifacts to be produced in the various process stages, necessitating rigorous screening, through which only the essential ones are adopted.

One of the major drawbacks of RUP is, however, that it fails to provide any clear implementation guidelines in the vein of, for example, Crystal, which lists the required documentation and roles with respect to software criticality and project size. RUP leaves the tailoring to the user entirely, which raises the question how much of the RUP can be left out, with the resulting process still being "RUP"?

### 3.5.6. Current research

dX, by Robert Martin, is a minimal version of RUP, in which agile software development viewpoints are taken into consideration, stripping the RUP activities down to their bare essentials (Martin 1998). In order to point out the malleability of RUP, dX deliberately mimics the principles of XP, but as tailored RUP activities (apart from meaning "a small change", "dX" is "XP" upside down).

Agile modeling (Ambler 2002a) is a new approach (see details in 3.9.1), which in RUP can be used for Business modeling, defining requirements, and analysis and design phases.

## 3.6. Dynamic Systems Development Method

Since its origin in 1994, DSDM, the Dynamic Systems Development Method, has gradually become the number one framework for rapid application development (RAD) in the UK. (Stapleton 1997). DSDM is a non-profit and non-proprietary framework for RAD development, maintained by the DSDM Consortium[10]. The developers of the method maintain that in addition to serving as a method in the generally accepted sense DSDM also provides a framework of controls for RAD, supplemented with guidance on how to efficiently use those controls.

The fundamental idea behind DSDM is that instead of fixing the amount of functionality in a product, and then adjusting time and resources to reach that functionality, it is preferred to fix time and resources, and then adjust the amount of functionality accordingly.

### 3.6.1. Process

DSDM consists of five phases: feasibility study, business study, functional model iteration, design and build iteration, and implementation (Figure 11). The first two phases are sequential, and done only once. The three last phases, during which the actual development work is done, are iterative and incremental. DSDM approaches iterations as timeboxes. A timebox lasts for a predefined period of time, and the iteration has to end within the timebox. The time allowed for each iteration to take is planned beforehand, along with the results the iteration is guaranteed to produce. In DSDM, a typical timebox duration is from a few days to a few weeks.

[10] See www.dsdm.org.

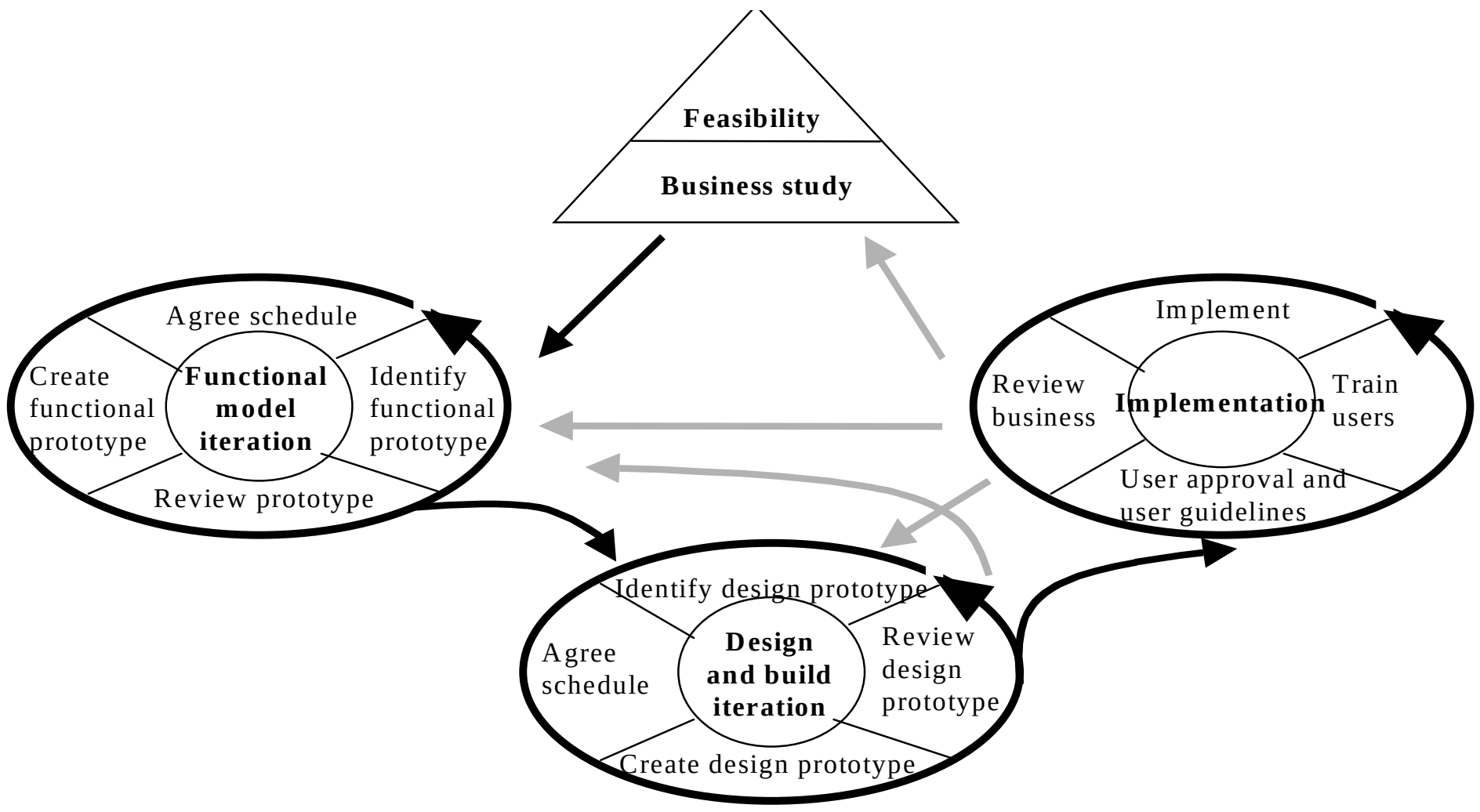

*Figure 11. DSDM process diagram (Stapleton 1997, p. 3)*

In the following, the phases are introduced, with their essential output documents.

The **feasibility study** phase is where the suitability of DSDM for the given project is assessed. Judging by the type of project and, most of all, organizational and people issues, the decision is made, whether to use DSDM or not. In addition, the feasibility study phase is concerned with the technical possibilities of going through with the project, and the risks therein. Two work products are prepared – a feasibility report, and an outline plan for development. Optionally, a fast prototype can also be made if the business or technology are not known well enough to be able to decide whether to proceed to the next phase or not. The feasibility study phase is not expected to take more than a few weeks.

The **business study** is a phase where the essential characteristics of the business and technology are analyzed. The recommended approach is to organize workshops, where a sufficient number of the customer's experts are gathered to be able to consider all relevant facets of the system, and to be able to agree on development priorities. The affected business processes and user classes are described in a Business Area Definition. The identification of the affected user classes helps involving the customer, as the key people in the customer's organization can be recognized and involved at an early stage. High level descriptions of the processes are presented in the Business Area Definition, in a suitable format (ER diagrams, business object models, etc.).

Another two outputs are done in the business study phase. One is a System Architecture Definition, and the other an Outline Prototyping Plan. The architecture definition is the first system architecture sketch, and it is allowed to change during the course of the DSDM project. The prototyping plan should state the prototyping strategy for the following stages, and a plan for configuration management.

The **functional model iteration** phase is the first iterative and incremental phase. In every iteration, the contents and approach for the iteration are planned, the iteration gone through, and the results analyzed for further iterations. Both analysis and coding are done; prototypes are built, and the experiences gained from them are used in improving the analysis models. The prototypes are not to be entirely discarded, but gradually steered towards such quality that they can be included in the final system. A Functional Model is produced as an output, containing the prototype code and the analysis models. Testing is also a continuing, essential part of this phase.

There are four further outputs in the phase (at different stages in the phase). **Prioritized functions** is a prioritized list of the functions that are delivered at the end of the iteration. **Functional prototyping review documents** collect the users’ comments about the current increment, working as input for subsequent iterations. **Non-functional requirements** are listed, mainly to be dealt with in the next phase. **Risk analysis of further development** is an important document in the functional model iteration phase, because from the next phase (design and build iteration) onwards, encountered problems will be more difficult to address.

The **design and build iteration** is where the system is mainly built. The output is a Tested System that fulfils at least the minimum agreed set of requirements. Design and build are iterative, and the design and functional prototypes are reviewed by the users, and further development is based on the users’ comments.

The final **implementation** phase is where the system is transferred from the development environment into the actual production environment. Training is given to users, and the system handed over to them. If the roll-out concerns a wide number of users, and done over a period of time, the implementation phase may also be iterated. Apart from the delivered system, the output of the implementation phase also includes a User Manual, and a Project Review Report. The latter summarizes the outcome of the project, and based on the results, the course of further development is set. DSDM defines four possible courses of development. If the system fulfils all requirements, no further work is needed. On the other hand, if a substantial amount of requirements have to be left aside (for example, if they were not discovered until the system was in development), the process may be run through again, from start to finish. If some less-critical functionality has to be omitted, the process may be run again from the functional model iteration phase onwards. Lastly, if some technical issues can not be addressed due to time constraints, they may be now done by iterating again, starting from the design and build iteration phase.

### 3.6.2. Roles and responsibilities

DSDM defines 15 roles for users and developers. The most dominant ones, as described in (Stapleton 1997) are listed in the following.

**Developers** and **senior developers** are the only development roles. Seniority is based on experience in the tasks the developer performs. The senior developer title also indicates a level of leadership in the team. The developer and senior developer roles cover all development staff, be it analysts, designers, programmers or testers.

A **Technical coordinator** defines the system architecture and is responsible for technical quality in the project. He is also responsible for technical project control, such as the use of software configuration management.

Of the user roles, the most important one is the **Ambassador User**. The respective duties are to bring the knowledge of the user community into the project, and to disseminate information about the progress of the project to other users. This ensures that an adequate amount of user feedback is received. The ambassador user has to come from the user community that will eventually use the system. Since the ambassador user is unlikely to represent all the required user viewpoints, however, an additional role of **Adviser User** is defined. Adviser users can be any users who represent an important viewpoint from the point of view of the project. Adviser users can be, e.g. IT staff, or financial auditors.

A **Visionary** is the user participant who has the most accurate perception of the business objectives of the system and the project. The Visionary is probably also the one with the initial the idea of building the system. The task of the visionary is to ensure that essential requirements are found early on, and that the project keeps going in the right direction from the viewpoint of those requirements.

An **Executive Sponsor** is the person from the user organization who has the related financial authority and responsibility. The Executive Sponsor therefore has ultimate power in making decisions.

### 3.6.3. Practices

Nine practices define the ideology and the basis for all activity in DSDM. The practices, called principles in DSDM, are listed in Table 4.

*Table 4. DSDM practices (Stapleton 1997).*

| **DSDM practice** | **Description** |
|---|---|
| Active user involvement is imperative. | A few knowledgeable users have to be present throughout the development of the system to ensure timely and accurate feedback. |
| DSDM teams must be empowered to make decisions. | Long decision-making processes cannot be tolerated in rapid development cycles. The users involved in development have the knowledge to tell where the system should go. |
| The focus is on frequent delivery of products. | Erroneous decisions can be corrected, if the delivery cycle is short and the users can provide accurate feedback. |
| Fitness for business | “Build the right product before you build it right”. Before the |

| DSDM practice | Description |
|---|---|
| purpose is the essential criterion for acceptance of deliverables. | core business needs of the system are satisfied, technical excellence in less important areas shouldn't be striven after. |
| Iterative and incremental development is necessary to converge on an accurate business solution. | System requirements seldom remain intact from the start of a project to its finish. By letting systems evolve through iterative development, errors can be found and corrected early. |
| All changes during development are reversible. | In the course of development, a wrong path may easily be taken. By using short iterations and ensuring that previous states in development can be reverted to, the wrong path can safely be corrected. |
| Requirements are baselined at a high level. | Freezing the core requirements should only be done at a high level, to allow the detailed requirements to change as needed. This ensures that the essential requirements are captured at an early stage, but development is allowed to begin before every requirement has been frozen. As the development progresses, more of the requirements should be frozen as they become clear and are agreed upon. |
| Testing is integrated throughout the lifecycle. | With tight time constraint, testing tends to be neglected if left to the end of the project. Therefore, every system component should be tested by the developers and users as they are developed. Testing is also incremental, and the tests built more comprehensive throughout the project. Regression testing is particularly emphasized because of the evolutionary development style. |
| A collaborative and cooperative approach shared by all stakeholders | In order for the DSDM to work, the organization must commit to it, and its business and IT departments cooperate. The choice of what is delivered in the system and what is left out is always a |

| DSDM practice | Description |
|---|---|
| is essential. | compromise, and requires common agreement.<br><br>On a smaller scale, the responsibilities of the system are shared, so the user / developer collaboration must also work seamlessly. |

### 3.6.4. Adoption and experiences

DSDM has been in widespread use in the UK since the mid 90’s. Eight case studies are documented in (Stapleton 1997), and the experiences clearly show that DSDM is a viable alternative for RAD.

In order to facilitate the method adoption, the DSDM Consortium has published a method suitability filter, in which three areas are covered: business, systems and technical. The main questions that have to be considered are listed in Table 5.

*Table 5. Adoption considerations (Stapleton 1997, pp. 19-22).*

| Question | Description |
|---|---|
| Is functionality going to be reasonably visible at the user interface? | If user involvement is attempted through prototyping, users need to be able to verify that the software is producing the correct actions without the user having to understand the technical details. |
| Can you clearly identify all classes of end users? | It is important to be able to identify and involve all relevant user groups of the software product. |
| Is the application computationally complex? | There is apparent danger in developing complex functionality from the scratch through the use of DSDM. Better results are obtained if some of the building blocks are already available for the |

| **Question** | **Description** |
|---|---|
| | development team. |
| Is the application potentially large? If it is, can it be split into smaller functional components? | DSDM had been used for the development of very large systems. Some of these projects have had a development period of 2 to 3 years. However, the functionality is to be developed in fairly discrete building blocks. |
| Is the project really time constrained? | Active user involvement is of paramount importance in the DSDM method. If the users do not participate in the development project (due to the fact that the project actually is not time constrained), developers may get frustrated and begin to make assumptions about what is needed in order to make the deadlines. |
| Are the requirements flexible and only specified at a high level? | If the detailed requirements have been agreed upon and fixed before the software developers begin their work, the benefits of DSDM will not be achieved. |

These and other adoption considerations can be found in the DSDM manual (DSDMConsortium 1997).

### 3.6.5. Scope of use

The DSDM team size varies between two and six, and there may be many teams in a project. The minimum of two persons involved comes from the fact that each team has to have at least one user and one developer. The maximum of six is a value found in practice. DSDM has been applied in small and large projects alike. The precondition for using it in large systems, as stated earlier, is that the system can be split into components that can be developed in small teams.

When considering the application domain, Stapleton (1997) suggests that DSDM is more easily applied to business systems than to engineering or scientific applications.

### 3.6.6. Current research

DSDM was originally developed and continues to be maintained by a consortium that consists of several member companies. DSDM manuals and supporting white papers are made available to consortium partners for a nominal annual cost. Outside the consortium there is no identifiable research to be found, while within the consortium the method continues to evolve. As an example, in 2001 an e-DSDM version of the method tailored for eBusiness and eCommerce projects was released (Highsmith 2002).

## 3.7. Adaptive Software Development

Adaptive Software Development, or ASD for short, was developed by James A. Highsmith III, and published in (Highsmith 2000). Many of ASD's principles stem from Highsmith's earlier research on iterative development methods. The most notable ancestor of ASD, "RADical Software Development", was co-authored by Highsmith and S. Bayer, and presented in (Bayer and Highsmith 1994).

ASD focuses mainly on the problems in developing complex, large systems. The method strongly encourages incremental, iterative development, with constant prototyping. Fundamentally, ASD is about "balancing on the edge of chaos" – its aim is to provide a framework with enough guidance to prevent projects from falling into chaos, but not too much, which could suppress emergence and creativity.

### 3.7.1. Process

An Adaptive Software Development project is carried out in three-phase cycles. The phases of the cycles are Speculate, Collaborate, and Learn (see Figure 12).

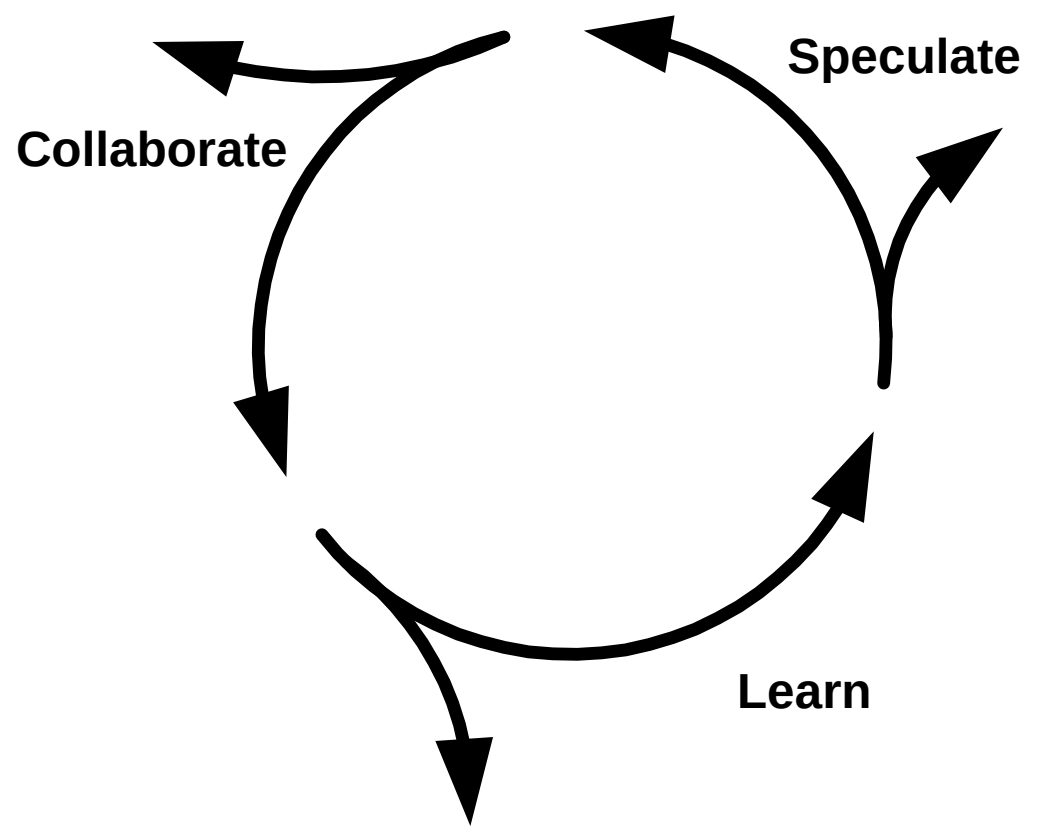

*Figure 12. The ASD cycle (Highsmith 2000, p. 18).*

The phases are named in a way to emphasize the role of change in the process. “Speculation” is used instead of “Planning”, as a “plan” is generally seen as something where uncertainty is a weakness, and from which deviations indicate failure. Similarly, “Collaborate” highlights the importance of teamwork as the means of developing high-change systems. “Learn” stresses the need to acknowledge and react to mistakes, and the fact that requirements may well change during development.

Figure 13 illustrates the adaptive development cycles in more detail. The Project Initiation phase defines the cornerstones of the project, and is begun by defining the project mission. The mission basically sets a coarse frame for the end product, and all development is steered so that the mission will be accomplished. One of the most important things in defining the project mission is to figure out what information is needed in order to carry out the project. The important facets of the mission are defined in three items: a project vision charter, a project data sheet, and a product specification outline. The meaning of these documents is explained in detail in (Highsmith 2000). The Project Initiation phase fixes the overall schedule, as well as the schedules and objectives for the development cycles. The cycles typically last between four and eight weeks.

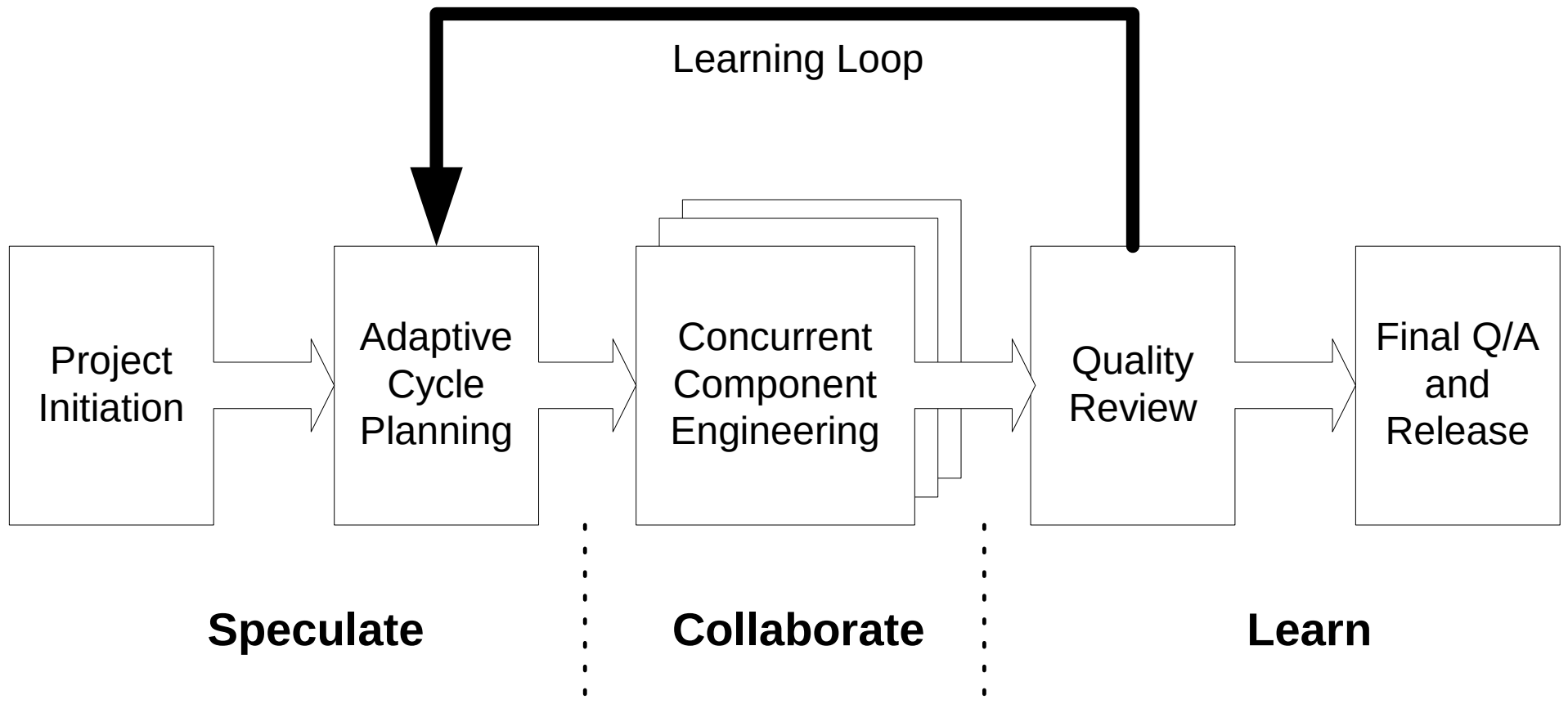

*Figure 13. The ASD lifecycle phases (Highsmith 2000, p. 84).*

ASD is explicitly component-oriented rather than task-oriented. In practice, this means that the focus is more on results and their quality rather than the tasks or the process used for producing the result. The way ASD addresses this viewpoint is through adaptive development cycles that contain the Collaborate phase, where several components may be under concurrent development. Planning the cycles is a part of the iterative process, as the definitions of the components are continuously refined to reflect any new information, and to cater for changes in component requirements.

Basis for further cycles (the “Learning Loop” in Figure 13) is gained from repeated quality reviews, which focus on demonstrating the functionality of the software developed during the cycle. An important factor in performing the reviews is the presence of the customer, as a group of experts (called customer focus-group). However, since the quality reviews are rather scarce (they take place only at the end of each cycle), customer presence in ASD is backed up by joint application development (JAD) sessions. A JAD session is essentially a workshop, where developers and customer representatives meet to discuss desired product features, and to enhance communication. ASD does not propose schedules for holding JAD sessions, but they are pointed out as particularly important in the beginning of a project.

The final stage in an Adaptive Software Development project is the Final Q/A and Release stage. ASD does not consider how the phase should be carried out, but stresses the importance of capturing the lessons learned. Project postmortems are seen as critically important in high-speed and high-change projects, where agile development takes place.

As a summary, the adaptive nature of development in ASD is characterized by the following properties:

*Table 6. Characteristics of adaptive development cycles* (Highsmith 2000).

| **Characteristic** | **Description** |
|---|---|
| Mission-Driven | The activities in each development cycle must be justified against the overall project mission. The mission may be adjusted, as the development proceeds. |
| Component-Based | Development activities should not be task-oriented, but rather focus on developing working software, i.e., building the system a small piece at a time. |
| Iterative | A serial waterfall method works only in well-understood and well-defined environments. Most development is turbulent, and the development effort should therefore be focused on *redoing* instead of *doing it right the first time*. |
| Time-Boxed | Ambiguity in complex software projects can be alleviated by fixing tangible deadlines on a regular basis. Time-boxed project management forces the project participants to make the inevitable, hard trade-off decisions early in the project. |
| Change-Tolerant | Changes are frequent in software development. Therefore, it is more important to be able to adapt to them, than it is to control them. To build a change-tolerant system, the developers must constantly evaluate whether the components they are building are likely to change. |

| Characteristic | Description |
|---|---|
| Risk-Driven | The development of high-risk items (e.g., least well known, or most critical if changed) should begin as early as possible. |

### 3.7.2. Roles and responsibilities

The ASD process largely originates from organizational and management culture and, especially, the importance of collaborating teams and teamwork. All these issues are given a lot of thought in (Highsmith 2000). The approach does not, however, describe team structures in detail. Likewise, very few roles or responsibilities are listed. An "execute sponsor" is named as the person with overall responsibility for the product being developed. Participants in a joint application development session are the only other roles mentioned (a facilitator to plan and lead the session, a scribe to make minutes, the project manager, and customer and developer representatives).

### 3.7.3. Practices

ASD proposes very few practices for day-to-day software development work. Basically, (Highsmith 2002) expressly names three: iterative development, feature-based (component-based) planning, and customer focus group reviews. Indeed, perhaps the most significant problem with ASD is that its practices are difficult to identify, and leave many details open. ASD has hardly any practices that *should* be done; most issues covered in the book are rather like examples of what *could* be done.

### 3.7.4. Adoption and experiences

The principles and the underlying ideas behind ASD are reasonable, but few guidelines are given for putting the method into use. The book (Highsmith 2000) offers a number of insights into software development in general, and makes a very interesting read. The philosophy behind the book is largely about building an adaptive organizational culture, not about the specifics. As a consequence, the method offers little basis for actually adopting ASD in an organization. Certainly, other methods have to be used along with it. This does not, however, imply that ASD is useless. Quite on the contrary – in an undisputed way, ASD provides important philosophy behind the agile movement, with insights on developing complex systems, collaboration, and agility in management culture.

### 3.7.5. Scope of use

ASD does not have built-in limitations for its application. One interesting feature in the scope of using ASD is that it does not enforce the use of co-located teams like most other agile software development methods. Building complex systems is seen to require extensive knowledge of many technologies, making it essential to bring together the expertise of many specialized teams.

Often the teams have to work across space, time and organizational boundaries. Highsmith (2000) makes note of this, and points out that most of the difficulties in distributed development are related to social, cultural and team skills. Hence, ASD offers techniques for enhancing inter-team collaboration by suggesting information sharing strategies, use of communication tools, and ways of gradually introducing rigor in project work in order to support distributed development.

### 3.7.6. Current research

There has not been significant research on ASD. Perhaps due to the fact that the method leaves much room for adaptation, few (if any) experiences have been reported of using ASD as is. The underlying principles of adaptive software development are not waning, however. The originator of ASD is taking the themes further, focusing on the central constituents of agile software development environments – namely, people, relationships and uncertainty. He has documented some of that work in a recent book, *Agile Software Development Ecosystems* (Highsmith 2002).

## 3.8. Open Source Software development

The discussion of Open Source Software (OSS) development is included in this chapter, and as noted in the previous chapter, it has several similarities with other (agile) software development methods, even though it also has its own peculiarities. The combination of the invention of the bulletin board system and the old custom of software developers to share code freely among colleagues was intensified and made possible in global scale with the expansion of the Internet in the '90s. This development process inspired a novel software development paradigm of OSS offering an innovative way to produce applications. This has aroused growing interest along with ample research efforts and discussions concerning the possibilities and meaning of the OSS paradigm for the whole software producing industry. This interest has increased notably after several success stories; among these are the Apache server, the Perl programming language, the SendMail mail handler, and the Linux operating system. Microsoft has even pronounced the later as their toughest competitor in the server operating systems market.

The OSS paradigm suggests the source code to be freely available for modifications and redistribution without any charges. The OSS paradigm can also be discussed from a philosophical perspective (O'Reilly 1999; Hightower and Lesiecki 2002). Feller and Fitzgerald (2000) present the following motivations and drivers for OSS development:

1) Technological; the need for robust code, faster development cycles, higher standards of quality, reliability and stability, and more open standards and platforms

2) Economical; the corporate need for shared cost and shared risk

3) Socio-political; scratching a developer's "personal itch", peer reputation, desire for "meaningful" work, and community oriented idealism.

Most known OSS development projects are focused on development tools or other platforms that are used by professionals who have often participated in the development effort themselves, thus having the role of the customer and that of the developer at the same time. OSS is not a compilation of well defined and published software development practices constituting a written eloquent method. Instead, it is better described in terms of different licenses for software distribution and as a collaborative way of widely dispersed individuals to produce software with small and frequent increments. The Open Source Initiative[11] keeps track of and grants licenses for software that complies with the OSS definitions. Several researchers, e.g. (Gallivan 2001; Crowston and Scozzi 2002), suggest that the OSS project represents in fact a virtual organization.

### 3.8.1. Process

Even though Cockburn (2002a) notes that OSS development differs from the agile development mode in philosophical, economical, and team structural aspects, OSS does in many ways follow the same lines of thought and practices as other agile methods. For example, the OSS development process starts with early and frequent releases, and it lacks many of the traditional mechanisms used for coordinating software development with plans, system level designs, schedules and defined processes. Typically, the OSS project consists of the following visible phases (Sharma et al. 2002):

1. Problem discovery
2. Finding volunteers
3. Solution identification
4. Code development and testing
5. Code change review
6. Code commit and documentation

[11] www.opensource.org, (13.5.2002)

7. Release management

Even though it is possible to depict the OSS software development methods with the above iteration stages, the interest lies in how this process is managed, as can be seen how the OSS development method is characterized by Mockus et al. (2000) with the following statements:

- The systems are built by potentially large numbers of volunteers
- Work is not assigned; people themselves choose the task they are interested in
- There exists no explicit system level design
- There is no project plan, schedule or list of deliverables
- The system is augmented with small increments
- Programs are tested frequently

According to Feller and Fitzgerald (2000), the OSS development process is organized as a massive parallel development and debugging effort. This includes loosely-centralized, cooperative and free of charge contributions from individual developers. However, there are also signs that this “free” development idea is changing. The OSS development process does not include any predefined formal documentation norms or customs. However, the process involves customs and taboos that are learned and perceived by experience.

### 3.8.2. Roles and responsibilities

The OSS development process seems to be quite free and wild. However, it must have some structure, or else it would never have been able to achieve such remarkable results as it has in the past years. Also old and established companies have started to show interest in OSS, among these notably IBM, Apple, Oracle, Corel, Netscape, Intel and Ericsson (Feller and Fitzgerald 2000). In these efforts, especially in larger OSS projects, the named companies have taken the role of the coordinator and that of the mediator, thus acting as project management.

Another new phenomenon is the emergence of the "auction" or repository sites like Freshmeat[12] and SourceForge[13], which have established development websites with a large repository of Open Source code and applications available on the Internet. These sites commonly provide free services to Open Source developers, including project hosting, version control, bug and issue tracking, project management, backups and archives, along with communication and collaboration resources. As no face-to-face meetings are used in the OSS context, the importance of the Internet is obvious. The development process inevitably needs well functioning versioning software that strictly tracks and allows the submission and prompt testing of new source code. The role of these tools cannot be underestimated, as the developed software must be orchestrated and run continuously, day and night, and the developers themselves have highly varying skill levels and backgrounds.

A typical OSS development effort structure, according to Gallivan (2001), is composed of several levels of volunteers:

1) Project leaders who have the overall responsibility for the project and who usually have written the initial code

2) Volunteer developers who create and submit code for the project. These can be further specified as:

    a. Senior members or core developers with broader overall authority

    b. Peripheral developers producing and submitting code changes

    c. Occasional contributors

---

[12] freashmeat.net, 20.05.2002

[13] sourceforge.net, 20.05.2002

d. Maintainers and credited developers

3) Other individuals who, in the course of using the software, perform testing, identify bugs and deliver software problem reports

4) Posters participate frequently in newsgroups and discussions, but do no coding.

Sharma et al. (2002) state that OSS development projects are usually divided by the main architects or designers into smaller and more easily manageable tasks, which are further handled by individuals or groups. Volunteer developers are divided in individual or small groups. They select freely the tasks they wish to accomplish. Thus the rational modular division of the overall project is essential to enable a successful outcome of the development process. Furthermore, these sub-tasks must be interesting so as to attract developers.

### 3.8.3. Practices

To start or to acquire an ownership of an OSS project can be done in several ways: to found a new one, to have it handed over by the former owner, or to voluntarily take over an ongoing dying project (Bergquist and Ljungberg 2001). Often the actual to-be users themselves define the product (project) and they also do the coding as well. The process is continuous, as software development is evolving endlessly. Even though hundreds of volunteers may be participating in the OSS development projects, usually there is only a small group of developers performing the main part of the work (Ghosh and Prakash 2000; Mockus et al. 2000; Dempsey et al. 2002).

Sharma and the others (2002) describe some of the central organizational aspects in the OSS approach, e.g. the division of labor, co-ordination mechanism, distribution of decision-making, organizational boundaries, and informal structure of the project. Mockus and the others (2000) characterize the OSS development as follows:

- OSS systems are built by potentially large numbers of volunteers

- Work is not assigned; people undertake the work they choose to undertake

- There is no explicit system-level design, or even detailed design

- There is no project plan, schedule, or list of deliverables.

To work successfully, the geographically dispersed individuals as well as small groups of developers must have well functioning and open communication channels between each other, especially as the developers do not usually meet face-to-face.

### 3.8.4. Adoption and experiences

Lerner and Tirole (2001) report in brief some experiences concerning the development paths of Linux, Perl, and Sendmail software, describing the development of these applications from an effort of a single programmer to a globally scattered team of individuals that can be counted in tens of thousands.

One of the problems of the OSS development process, and at the same time one of its strengths, is that OSS requirements are constantly being elaborated, and that the product is never in a finalized state, until all development work has ceased on the product (Feller and Fitzgerald 2000). The OSS development method is highly dependent on the Internet enabling global networked collaboration.

Another central question regarding OSS software is the stabilizing and commercializing of the code to an application level. As the OSS code itself is not a commercial product, a completely new service business has emerged focusing on producing commercially successful products out of these open codes. These new service businesses have taken the role of producing a packaged application with all the frills it needs to be commercially viable. Examples of such are, e.g., the Redhat and SOT for Linux. Software developed under the OSS paradigm can also be used in new business endeavors, in which case the whole business idea is founded on using only OSS-based software (Wall 2001).

### 3.8.5. Scope of use

At present, the most widely published results concentrate mostly on development efforts concerning large software tools and Internet based products (Lerner and Tirole 2001). The OSS development method itself does not bring forth any special application domains nor does it set any limits for the size of the software, as long as the suggested development projects comply with the elements that the OSS development methods are founded on as described above. Future trends also rely on taking advantage of global developer collaboration, while OSS development efforts are continuously producing raw material that can later be harnessed to commercial exploitation (O'Reilly 1999). From a legal perspective, the OSS development paradigm should be seen more as a licensing structure exploiting the terms of the General Public License. Typical examples of OSS applications are, according to Feller (2000), complex back-office infrastructural support systems with a high number of users.

The OSS development process itself can be implemented with different specific software development methods, although these methods must comply with and allow for the characteristics of the OSS paradigm. Thus the process can have agile aspects, but on the other hand, the development process itself can be slowed down resulting from several reasons, e.g., lack of interested and skillful developers, and conflicting views.

Geographically and culturally dispersed organizations could benefit from analyzing the pros and cons of the different OSS methods and taking the good ones into use in their specific context of software development.

### 3.8.6. Current research

Numerous research efforts are being made to explicate the contradiction of the success of the OSS projects. How does it function, as there is no formal OSS method, but still the organic group of developers is able to produce usually well functioning software. One hot topic is the legal aspects that follow the use the OSS components. Feller and Fitzgerald (2000) have analyzed the OSS licensing practices and Lerner and Tirole (2001) treat briefly the question of optimal licensing in the OSS environment. They discuss the problems of free riding and “hijacking” the code developed in joined action for further proprietary commercial use.

Software companies are showing more and more interest in finding out the possibilities of harnessing the OSS methodology to support their daily work. Sharma and the others (2002), for example, have investigated this problem and suggested a framework for creating a hybrid-OSS community structure. Various companies are also studying the imminent effects of the new business models brought along by the OSS software production and distribution methods (O'Reilly 1999). Still, according to Sharma and the others (2002), only some empirical studies have been conducted and little rigorous research has been done on how traditional software organizations can implement OSS practices and benefit from them.

The network effect is crucial for the success of OSS development. New prototypes are introduced on the net. However, the community is harsh, very much like normal business world, i.e. if you find the customers you survive, but without customers you die. That is, if the introduced prototype gathers enough interest, then it will gradually start to attract more and more developers. In some cases OSS provides raw material that can be further developed and extended to comply with commercial needs. Bergquist and Ljungberg (Bergquist and Ljungberg 2001, p. 319) have studied the relationship development in OSS communities. They conclude that “one could understand this culture [OSS] as a kind of amalgamation of collectivism and individualism: giving to the community is what makes the individual a hero in the eyes of others. Heroes are important influences but also powerful leaders.”

Crowston and Scozzi (2002) have investigated several OSS projects using the data available at Sourceforge.net. Their main propositions are:

1. The availability of competencies is a key factor in the success of projects.

2. It is important for the developers to be able to recognize the needs of the customers of the project. Further, the projects related to systems development and administration were found to be more successful.

3. Projects manned with well-known administrators are more successful, as they attract new volunteers more easily.

## 3.9. Other Agile Methods

In the previous subsections, several methods for producing software in an agile manner have been presented. The focus has been on methods that provide comprehensive frameworks with more or less tangible processes and practices that cover as much of the software development process as possible.

The agile approach has, however, recently brought about a great deal of research and lots of interesting new ideas, which either have not been documented thoroughly (e.g., Lean Software Development[14]) or have been published just very recently (e.g., Agile Modeling). These types of methods will be at focus in this section since they arguably provide interesting viewpoints in different areas of agile development.

### 3.9.1. Agile Modeling

Agile Modeling, which was introduced by (Ambler 2002a) in 2002 is a new approach to performing modeling activities. The key focus in AM is on practices and cultural principles. The underlying idea is to encourage developers to produce advanced enough models to support acute design problems and documentation purposes, while still keeping the amount of models and documentation as low as possible. The main focus on cultural issues is reflected in the support for communication, team structures and the ways the teams work together.

Four of the values behind Agile Modeling are the same as in Extreme Programming: communication, simplicity, feedback, and courage. Fundamentally, these promote similar issues as in XP. In addition, Ambler cites humility, emphasizing the fact that different people have different sorts of expertise, which should be exploited by encouraging co-operation and getting the best people to do the given job.

[14] www.poppendieck.com/

The core principles behind Agile Modeling promote the importance of functioning software as the main goal of modeling. Not unlike XP, the principles partly arise from the grim reality that changes in software development are frequent, and have to be considered in modeling as well. The core principles are supported by supplementary principles helping to produce efficient models.

The gist of Agile Modeling is in its practices. While the author notes that each of the core principles have to be complied with in order for the modeling to be Agile Modeling, the practices listed in AM can be taken into use independently of each other. Agile Modeling is strongly focused on practices. Eleven best practices are listed as core practices in four categories (Iterative and Incremental Modeling, Teamwork, Simplicity, Validation). The practices range from the selection of a proper modeling technique for the item to be modeled, to the benefits of teamwork. Eight supplementary practices are also listed in three categories (productivity, documentation, motivation).

Agile Modeling is not a mere collection of practices, but contains also a lot of thought about the way the practices fit together and how they should be supported by issues of organizational culture, everyday working environments, tools and teamwork. Like many agile methods, Agile Modeling favors customer presence, small teams and close communication. No limits are given, however, to the size of the systems that could be built using Agile Modeling as the modeling principle.

From a broader software development viewpoint, Agile Modeling is not sufficient by itself. It requires supporting methods, as it covers only modeling – and even that viewpoint is deliberately limited so that the book does not describe how to create, for example, UML models. Since the aim of Agile Modeling is to support day-to-day software development duties, however, significant effort has been put to illustrating the possibilities of seamless integration of Agile Modeling with other development methodologies. Extreme Programming and Rational Unified Process are presented as illustrations of combining Agile Modeling with other development methodologies.

Agile Modeling successfully points out that many agile software development principles suit modeling equally well, and provides thought-out illustrations of how the agile principles impact modeling. However, Agile modeling is a new method, and no identifiable research is available of it yet.

### 3.9.2. Pragmatic Programming

The title is slightly misleading, as there is no method called "pragmatic programming". Instead, there is an interesting set of programming best practices, published in "The Pragmatic Programmer" by Andrew Hunt and David Thomas (2000). Here, for the sake of simplicity, the approach shall be called "pragmatic programming".

The reason for including pragmatic programming in this review is that both authors of the book have been involved in the agile movement, and were among the originators of the Agile Manifesto. More importantly, the techniques covered by the book very concretely augment the practices discussed in the other methods.

Pragmatic programming does not have a process, phases, distinct roles or work products. It does, however, cover most programming practicalities in a very undisputable, well-grounded way. Throughout the book, the authors stress their main points as short tips. Many of the tips (there are a total of 70 of them) focus on day-to-day problems, but the philosophy behind them all is fairly universal and can be applied to any software development phase.

The philosophy is defined in six points, which could be stated roughly as:

- Take responsibility for what you do. Think solutions instead of excuses.
- Don't put up with bad design or coding. Fix inconsistencies as you see them, or plan them to be fixed very soon.
- Take an active role in introducing change where you feel it is necessary.
- Make software that satisfies your customer, but know when to stop.

- Constantly broaden your knowledge.
- Improve your communication skills.

From an agile viewpoint, the most interesting practices focus on incremental, iterative development, rigorous testing, and user-centered design. The approach has a very practical standpoint, and most practices are illustrated by positive and negative examples, often complemented by questions and code snippets. Considerable effort is put in explaining how to design and implement software so that it can withstand changes. Solutions for keeping software consistent throughout the changes are also discussed, refactoring being one solution.

The approach of pragmatic programming to testing involves the test code being implemented alongside the actual code, and all tests being made automatic. The idea is that if each corrected bug is not added into the test library and if regression tests are not habitually run, time and effort is wasted in finding the same bugs over and over again, and the adverse effects of changes in code cannot be detected early enough.

Automatization is encouraged in pragmatic programming for many other tasks as well. In fact, Tip #61 in the book goes as far as to suggest "Don't use manual procedures". For example, creating documentation drafts from source code, creating code from database definitions and automating nightly builds are mentioned as things that should be automated. Examples of suitable software for putting automation into place are given.

Spanning slightly fewer pages than the practices discussed above, the book considers the importance of actively digging for requirements, and keeping specifications at a reasonable level of detail. The quality goal throughout these phases, and throughout the entire book, is to provide "Good enough software".

In a nutshell, pragmatic programming demonstrates simple, responsible and disciplined software practices. It illustrates many practical points that can be adopted regardless of the surrounding methods or processes. The practices are mostly written from the point of view of a programmer, ranging from how to avoid typical coding and design errors, to communication and teamwork issues.

While certainly not a catch-all source of information on running a software project, the book is a good source of practical tips that will probably benefit any programmer or designer.

# 4. Comparison of agile methods

The purpose of this chapter is to compare the agile software development methods introduced in Chapter 3. This chapter is organized as follows. First, the analysis method, which guides the comparison, is briefly introduced. This is followed by a comparison of the selected general features. Finally, some aspects of adopting the method are considered and some future research needs are identified.

## 4.1. Introduction

The task of objectively comparing just about any methodology with another is difficult and the result is often based upon the subjective experiences of the practitioner and the intuitions of the authors (Song and Osterweil 1991). Two alternative approaches exist: informal and quasiformal comparison (Song and Osterweil 1992). Quasiformal comparison attempts to overcome the subjective limitations of an informal comparison technique. According to Sol (1983) quasiformal comparisons can be approached in five different ways:

1. Describe an idealized method and evaluate other methods against it.

2. Distill a set of important features inductively from several methods and compare each method against it.

3. Formulate *a priori* hypotheses about the method's requirements and derive a framework from the empirical evidence in several methods.

4. Define a metalanguage as a communication vehicle and a frame of reference against which you describe many methods.

5. Use a contingency approach and try to relate the features of each method to specific problems.

Song and Osterweil (1992) argue that the second and fourth type of approach are closer to the classic scientific method used for method comparison purposes. In fact, the fourth approach, i.e. detailing a metalanguage as a communication vehicle, was operationalized in the previous chapter, in which each method was communicated using the same generic typology, i.e. process, roles and responsibilities, practices, adoption and experiences, scope of use and current research.

Comparison often implies valuing one method over another. This, however, is not the intent here. Rather, by using the second approach, some important features concerning the method and its adoption are chosen as perspectives through which the methods are analyzed. Our goal, then, is to identify the differences and similarities between different agile software development methods. Recently, a similar approach was used in the survey on software architecture analysis methods (Dobrica and Niemelä 2002).

## 4.2. General features

In connection with the analytical elements used in Chapter 3, general features refer to current research and scope of use. In addition, we shall employ terms "key points" and "special features". Key points detail the methods' principal aspects or solutions. Special features describe one or several aspects of the methods that differentiate them from the others.

DSDM and Scrum were introduced in the early and mid 1990's, respectively. However, when determining the method status Scrum still remains in a "building up" phase, since studies citing to the use of the method are still scarce. Other methods that are also widely recognized are FDD, Crystal, PP and ASD. However, less is known about their actual usage. Thus, they can also be considered to be in the "building up" phase. AM was proposed under a year ago and thus its status is "nascent". XP, RUP, OSS and DSDM, on the other hand, are methods or approaches that have been well documented and of which there are a number of literature and experience reports available. Furthermore, each method has generated its own active research and user communities. Thus, they can be classified as "active".

None of the methods can be classified as "fading", even though, for example, the use of DSDM terminology, such as prototyping, is considered to be outdated. In actual use, prototyping could be referred to by the agile definition of working

code(Highsmith 2002). Performance and capacity prototypes can then be mapped to the XP spike, which is a piece of test code for a specific system aspect that may be thrown away. The status of agile software development methods is summarized in Table 7.

*Table 7. Agile SW development method status*

| Status (8/02) | Description | Method |
|---|---|---|
| Nascent | Method has been available less than one year. No identifiable research exists, no experiences reported. | AM |
| Building up | Method or approach is recognized widely, experience reports showing up, active community building up, research regarding the method identifiable. | FDD, Crystal, Scrum, PP, ASD |
| Active | Method detailed in several places, experience reports, active research, active community. | xP, RUP, OSS, DSDM[15] |

As stated in chapter 2, all agile methods (expect for RUP and OSS) are based on a chaordic perspective, they all share a similar set of collaborative values and principles and they value having a barely sufficient methodology (Highsmith 2002). However, each method is approaching the problems faced in software engineering from a different angle. Table 8 summarizes each method using three selected aspects: key points, special features and identified shortcomings. As stated, key points detail the methods’ principal aspects or solutions and special features describe one or several aspects of the methods that differentiate them from the others. Finally, identified shortcomings relate to some aspects of a method that are clearly missing or other aspects that have been documented in the literature.

[15] DSDM is developed and used mainly by the members of the DSDM Consortium. The method book is, however, publicly available.

*Table 8. General features of agile software development methods.*

| Method name | Key points | Special features | Identified shortcomings |
|---|---|---|---|
| ASD | Adaptive culture, collaboration, mission-driven component based iterative development | Organizations are seen as adaptive systems. Creating an emergent order out of a web of interconnected individuals. | ASD is more about concepts and culture than the software practice. |
| AM | Applying agile principles to modeling: Agile culture, work organization to support communication, simplicity. | Agile thinking applies to modeling also. | This is a good add-on philosophy for modeling professionals. However, it only works within other methods. |
| Crystal | Family of methods. Each has the same underlying core values and principles. Techniques, roles, tools and standards vary. | Method design principles. Ability to select the most suitable method based on project size and criticality | Too early to estimate: Only two of four suggested methods exist. |
| DSDM | Application of controls to RAD, use of timeboxing, empowered DSDM teams, active consortium to steer | First truly agile software development method, use of prototyping, several user roles: "ambassador", | While the method is available, only consortium members have access to white papers dealing with the actual use of the |

| Method name | Key points | Special features | Identified shortcomings |
|---|---|---|---|
| | the method development. | “visionary” and “advisor”. | method. |
| XP | Customer driven development, small teams, daily builds | Refactoring – the ongoing redesign of the system to improve its performance and responsiveness to change. | While individual practices are suitable for many situations, overall view & management practices are given less attention. |
| FDD | Five-step process, object-oriented component (i.e., feature) based development. Very short iterations: from hours to 2 weeks. | Method simplicity, design and implement the system by features, object modeling. | FDD focuses only on design and implementation. Needs other supporting approaches. |
| OSS | Volunteer based, distributed development, often the problem domain is more of a challenge than a commercial undertaking. | Licensing practice; source code freely available to all parties. | OSS is not a method itself; ability to transform the OSS community principles to commercial software development. |
| PP | Emphasis on pragmatism, theory of programming is of less importance, high level of automation in all aspects of | Concrete and empirically validated tips and hints, i.e., a pragmatic approach to software | PP focuses on important individual practices. However, it is not a method through which a system can be |

| Method name | Key points | Special features | Identified shortcomings |
|---|---|---|---|
| | programming. | development. | developed. |
| RUP | Complete SW development model including tool support. Activity driven role assignment. | Business modeling, tool family support. | RUP has no limitations in the scope of use. A description how to tailor, in specific, to changing needs is missing. |
| Scrum | Independent, small, self-organizing development teams, 30-day release cycles. | Enforce a paradigm shift from the "defined and repeatable" to the "new product development view of Scrum." | While Scrum details in specific how to manage the 30-day release cycle, the integration and acceptance tests are not detailed. |

Table 8 shows differences in the key points the studied methods place emphasis on. ASD is the most abstract method from the software development viewpoint. Although very appealing, its key goal – "creating an emergent order out of a web of interconnected individuals" – may, however, be difficult to grasp. This is also its major shortcoming since practitioners may experience difficulties in translating the method's "new" concepts to their use. Agile modeling (AM), XP and pragmatic programming all represent practice-oriented viewpoints. They all contain a number of empirically validated practices found useful by practitioners. As such, they are very valuable. The Crystal family of methodologies is the only one to explicitly suggest method design principles to allow tailoring depending on project size and criticality. This is an important aspect since method scalability is one of the major topics that the agile community needs to address. Attempts in this direction can also be identified (see discussion in e.g. eWorkshop 2002).

DSDM is differentiated from the other methods by its use of prototyping. DSDM also puts forward some roles that others have not considered. Such user (or customer related) roles are ambassador, visionary and advisor. These user roles represent different customer viewpoints. The drawback of DSDM is the requirement to belong to the DSDM consortium in order to gain an access to the white papers discussing different aspects of the method. FDD does not attempt to provide any all-in-one solution to software development, but focuses into a simple five-step approach, which is based on identifying, designing and implementing features. FDD presupposes that some work towards the project has been done already. Consequently, it does not cover the early phases of a project (see next subsection for more details on the life-cycle support). Scrum is a project management approach that relies on self-organizing independent development teams implementing a software project in 30-day cycles called sprints. The method book is not very clear and, e.g., integration and acceptance tests are not described.

Similar to ASD, OSS is more of a development philosophy than a method *per se*. However, many successful projects have been reported in literature. A special feature of OSS is the licensing practice, which requires that the source code is made freely available to all parties. The source code can then be read, modified, compiled and then redistributed free of charge. Finally, RUP is not considered a particularly agile software development method, but it can be one. It stands out from the others by being a complete development suite supported by various commercially sold tools. This is something that is mostly missing from the other methods. RUP also extends the method to contain business modeling practices similar to DSDM. Thus, they provide support also to the early phases of a software development project.

## 4.3. Adoption

In connection with the analytical elements used in Chapter 3, adoption here refers to process, roles and responsibilities and adoption and experiences. In addition, the comparison will be extended to include also a project management point of view, which sheds light into understanding how a method supports the managerial perspective.

Software products yield business benefits only if they are used. Similarly, the benefits associated with agile software development methods are obtainable only if these methods are used in the production process. The adoption of a new process technology should be relatively easy to perform since organizations can not afford to slow or stop the production to reorganize or to learn new ways of doing their business (Krueger 2002). The adoption and experiences regarding each method were briefly considered in Chapter 3. It is rather evident that there are not many experience reports available. Scientific studies are even harder to find at this stage.

Nandhakumar and Avison (1999, p. 188) found that currently, traditional software development methods are used "as a necessary fiction to present an image of control or to provide a symbolic status." They suggest that "alternative approaches that recognize the particular character of work in such environments are required." The particular character of work they referred to is the turbulent business environment or "the change-driven economy" as the agilists call it (e.g., Highsmith 2002). In this environment, changes are expected to arrive into the project even at a very late phase. Nandhakumar and Avison (1999) found that the developers' work practices are best characterized as having an improvisatory character where team members are protective of their professional autonomy and social controls have more influence on their practice than the suggested methodologies. These are, again, issues that all agile software development methods address explicitly. In fact, all of the reviewed agile methods can be characterized as placing emphasis on the following aspects: 1) delivering something useful, 2) reliance on people, 3) encouraging collaboration, 4) promoting technical excellence, 5) doing the simplest thing possible and 6) being continuously adaptable (Highsmith 2002).

Nandhakumar and Avison (1999) argued that current SW development is very different from the popular methodological view (see also Chapter 2 for related discussion). Table 9 complements their findings with the agile approach perspective. Table 9 shows further that the agile software development approaches embody a lot of the current software development practice. Thus, this explains in part why, e.g., extreme programming has received attention in recent years. It explicitly approaches the software development from the developers' point of view.

*Table 9. Comparing software development views (Nandhakumar and Avison 1999), augmented with the agile methods viewpoint.*

| | Methodological view | Software Development in practice | A view advocated by the agile methods |
|---|---|---|---|
| Activities | Discrete tasks | Interrelated personal projects | Interrelated personal projects |
| | Duration predictable | Completion unpredictable | Completion of the next release predictable |
| | Repeatable | Context dependent | Often context dependent |
| Process Performance | Reliable | Depends on contextual conditions | Performance tied to a small release, thus reliable |
| | Specifiable interactions | Inherently interactive | Specified interactions promoting the inherently interactive nature |
| | Single tasks in sequence | Many tasks are interwoven | Single tasks often not mandated due to their context dependent nature |

| | Methodological view | Software Development in practice | A view advocated by the agile methods |
|---|---|---|---|
| Developers' efforts | Dedicated to SW projects | Common to all activities (e.g., project, non-project, personal, routines) | Developers estimate the work effort required |
| | Undifferentiated | Specific to individuals | Specific to individuals |
| | Fully available | Fully utilized | Fully utilized |
| Control of work | Regularity | Opportunism, improvisation and interruption | Only mutually agreed controls exist[16], respect for the work of others |
| | Progress milestones, planning and management control | Individual preference and mutual negotiation | Individual preference and mutual negotiation |

While agile approaches concur with the current software development practice, they are not all suitable for all phases in the software development life-cycle. Figure 14 shows which phases of software development are supported by different agile methods. Each method is divided into three blocks. The first block indicates whether a method provides support for project management. The second block identifies whether a process, which the method suggests to be used, is described within the method. The third block indicates whether the method describes the practices, activities and workproducts that could be followed and used under varying circumstances. A gray color in a block indicates that the method covers the life-cycle phase and a white color indicates that the method does not provide detailed information about one of the three areas evaluated, i.e. project management, process or practices / activities / work products.

[16] This is not the case regarding some agile methods. As an example, DSDM and Crystal define in detail the controls that are used.

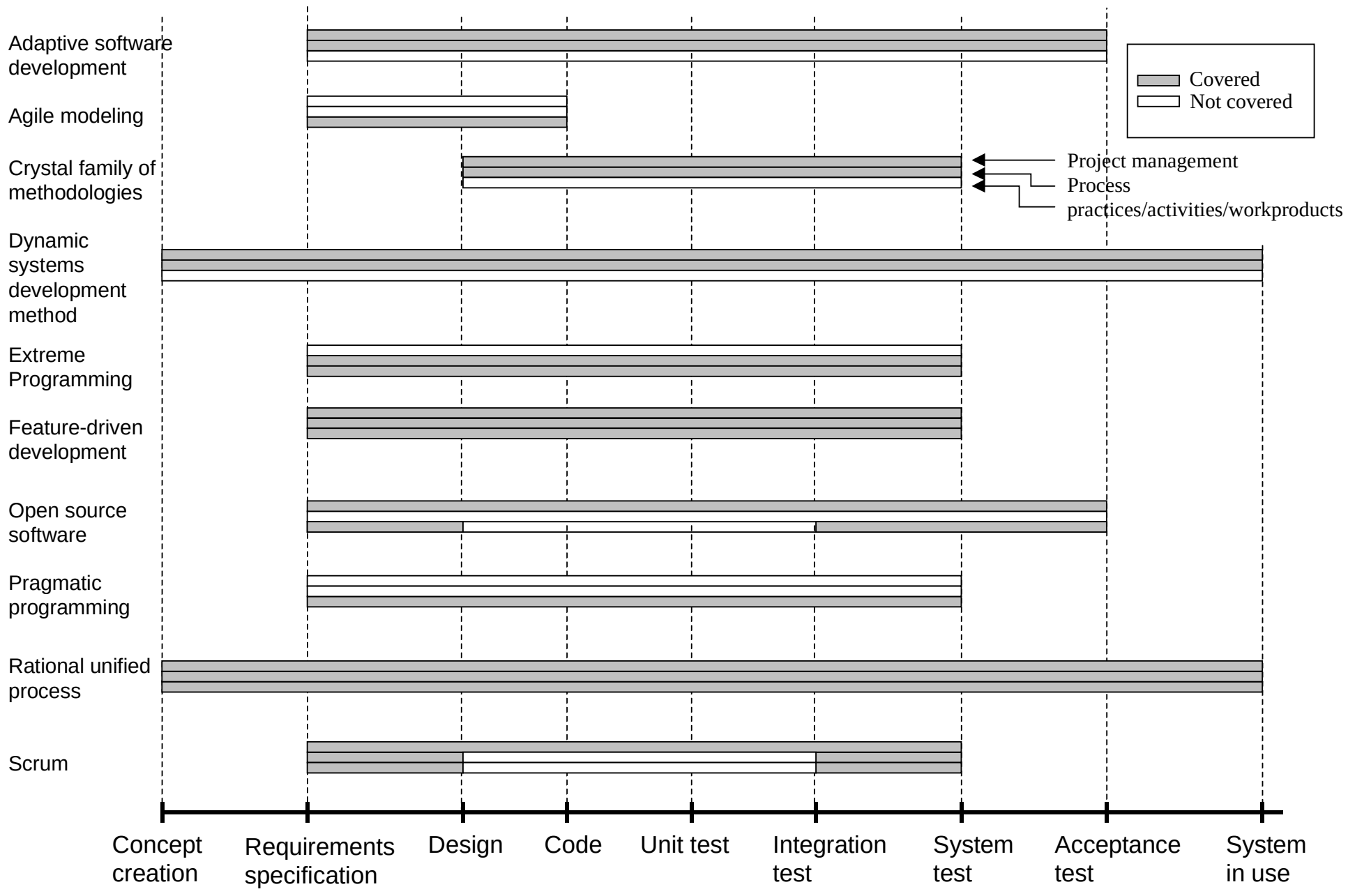

*Figure 14. Software development life-cycle support*

Figure 14 shows that agile methods are focused on different aspects of the software development life-cycle. Moreover, some are more focused on practices (extreme programming, pragmatic programming, agile modeling), while others focus on managing the software projects (Scrum). Yet, there are approaches providing full coverage over the development life cycle (DSDM, RUP), while most of them are suitable from the requirements specification phase on (e.g., FDD). Thus, there is a clear difference between the various agile software development methods in this regard. Whereas DSDM and RUP do not need complementing approaches to support software development, the others do to a varying degree. However, DSDM is available only for the members of the DSDM consortium, and RUP, then, is a commercially sold development environment.

Considering the adoption of the methods, the size of the development team is currently one of the main decisive issues. XP, Scrum, AM and PP are focused on small teams, preferably less than 10 software engineers. Crystal, FDD, RUP, OSS, ASD and DSDM claim to be capable of scaling up to 100 developers. However, agile proponents agree that when the development team size gets larger, the amount of, e.g., documentation is likely to (and should) increase, thus making the project "less agile". When the development group exceeds 20 software engineers, agilists place emphasis on solving the communication problems efficiently (Ambler 2002c). Each method contains a number of suggestions how to organize communication channels within a small group of engineers.

The agile approach appears particularly suitable in a situation where future requirements are not known (e.g., Ambler 2002c). It appears that if future requirements (performance requirements, exception handling, functionality etc.) are known and can be specified, agile approaches do not provide added value to a development project *per se*. Highsmith (2002) made an attempt to work out which market phase would be the most suitable for agile methods. He suggests that agile approaches are particularly suitable for "bowling alley" and "tornado" phases (see characterizations of different market phases in Moore 1995). As an example, in the bowling alley, Highsmith argues, "high levels of customization and intense customer interaction could be enhanced by a collaboration-oriented agile approach". Highsmith also suggests that matching the methodology type to the organization culture is important. Organizations that value competence and intense collaboration are more suitable for agile approaches than those relying on control or cultivation. No empirical support for these claims, however, is provided.

Adoption costs and difficulty are hard to evaluate since only a very limited number of studies have been published regarding these issues. However, we maintain that if the paradigm shift between the traditional way of software development and the agile software development is too large and supporting adoption models and techniques are missing, it is more than likely that organizations and developers will not be keen in adopting agile methods into everyday use. This paradigm shift includes placing importance on other issues than what traditionally have had focus such as planning, preparation for the future, documentation, contract negotiation, along with processes and tools. Organizations and people are not likely to make 180 degree turns in their software development practices.

Most of the methods studied call for empowering the customer and the development team to make important decisions regarding the development process. Thus, adopting agile SW practices also requires a cultural shift, especially in the middle and top management. Many of the traditional control mechanisms, such as periodic reporting, are shifted towards producing working code instead. On one hand, this requires changing the way the organization is functioning and, more importantly, on the other hand, placing more trust on the ability and competence of the development team. On the other end of the spectrum, there is the risk associated with the change of development approach. All of the agile approaches propose the development of software features in small incremental cycles, only 2 to 6 weeks in duration. Therefore, the risk is minimized since, e.g., a schedule slippage will be visible in a few weeks time.

Well-known agile methods proponents, such as Highsmith and Schwaber, believe that better results in method adoption are obtained when “the project has its’ back against the wall” (eWorkshop 2002; Highsmith 2002) thus ensuring managerial attention and support. While this type of anecdotal evidence is useful, empirically validated or purely experimental studies are needed to determine when, how and in which situations agile methods are best used and applied.

The adoption of new technology is a well-documented area of study (for more details and respective references, see e.g. Sultan and Chan 2000). It is less likely, however, that agile software development technologies are so radically different from other software technologies that would hinder the value of research in the area of innovation-diffusion. For example, Agarwal and Prasad (2000) show that beliefs about new technology and the subsequent adoption of the technology are strongly related. Abrahamsson (2002) made similar arguments. Other important factors affecting the adoption of new technology are organizational tenure, prior technical knowledge, training experiences, and perceived job insecurity. These findings, among others, should be used in the development of adoption models targeted to provide guidance in the process of introducing agile software development methods.

# 5. Conclusions

Studies have shown that traditional plan-driven software development methodologies are not used in practice. It has been argued that the traditional methodologies are too mechanistic to be used in detail. As a result, industrial software developers have become skeptical about "new" solutions that are difficult to grasp and thus remain unused. Agile software development methods, "officially" started with the publication of the agile manifesto, make an attempt to bring about a paradigm shift in the field of software engineering. Agile methods claim to place more emphasis on people, interaction, working software, customer collaboration, and change, rather than on processes, tools, contracts and plans. A number of promising new methodologies claiming conformance to these agile principles have been introduced. Yet, no systematic review of the agile methods has been attempted.

This publication had three purposes. Firstly, it synthesized existing literature on what is actually meant by the term "agile" by asking the question: "What makes a development method an agile one?" The conclusion was that this is the case when software development is

- incremental (small software releases, with rapid cycles),
- cooperative (customer and developers working constantly together with close communication),
- straightforward (the method itself is easy to learn and to modify, well documented), and
- adaptive (able to make last moment changes).

Secondly, based on this definition, each method was described in terms of process, roles and responsibilities, practices, adoption and experiences, along with current research. Thirdly, this enabled the selection of criteria for comparing methods and pointing out their differences and similarities. It was found that while the agile methods share many similarities, some are more focused than others. For example, they support different phases of the software product development to a varying degree. Differences were also found in the level of concreteness in which the methods are tied to actual development practice. For example, ASD is predominantly focused on principles that guide the software development while XP places emphasis on the development practices. These and other points were discussed in Chapter 4.

This, however, is only a starting point for our study. Anecdotal evidence suggests that agile methods are effective and suitable for many situations and environments. However, at present very few empirically validated studies can be found to support these claims. The existing evidence consists mainly of success stories from practicing professionals. Although these provide valuable information about practical applications, empirical studies are urgently needed for evaluating the effectiveness and the possibilities of using agile software development methods. Moreover, the frequent release of new agile methods is more likely to bring about confusion rather than clarity. Each method uses its own vocabulary and terminology and little work is being done in integrating the many viewpoints expressed. What is urgently needed now (more than new models) is the adoption or selection models to be used by practitioners. The goal of this type of research is to enable software professionals, projects and organizations to choose and to apply the right method at the right time. We believe that empirical studies or experiments provide a fruitful platform for this type of development.

In conclusion, agile thinking is a people-centric view to software development. People-centered strategies have been argued for as an important source of competitive advantage, because, unlike technology, cost, or new product development, these human strategies are difficult to imitate (Pfeffer 1998; Miller and Lee 2001). This, however, is not a new realization. A 1990 summer issue of the American Programmer (Ed Yourdon's Software Journal, Vol. 3, No. 7-8) was devoted exclusively to 'Peopleware'. The editor comments the special issue by pointing out that "Everyone knows the best way to improve software productivity and quality is to focus on people." Thus, the ideas that agile methods bring about are not new, nor do the agilists claim so. We, however, believe that agile software development methods provide a novel way of approaching software engineering problems, while also maintaining that the methods are by no means exhaustive or capable of solving all the problems. Majority of the software in large organizations is not produced by co-located teams of less than ten engineers. If, for example, scalability problems are not solved, agile thinking will not gain the serious attention it deserves.

# References

Abrahamsson, P. (2002). “Commitment nets in software process improvement.” Annals of Software Engineering, in press.

Agarwal, R. and J. Prasad (2000). “A field study of the adoption of software process innovations by information systems professionals.” IEEE Transactions on Engineering Management 47(3): 295-308.

Alexander, C. (1979). The Timeless Way of Building. New York, Oxford University Press.

Ambler, S. (2002a). Agile Modeling: Effective Practices for Extreme Programming and the Unified Process. New York, John Wiley & Sons, Inc. New York.

Ambler, S. (2002b). “Lessons in Agility from Internet-Based Development.” IEEE Software March / April: 66 - 73.

Ambler, S. W. (2002c). “Duking it out.” Software Development 10.

Anderson, A., R. Beattie, K. Beck, D. Bryant, M. DeArment, M. Fowler, M. Fronczak, R. Garzaniti, D. Gore, B. Hacker, C. Handrickson, R. Jeffries, D. Joppie, D. Kim, P. Kowalsky, D. Mueller, T. Murasky, R. Nutter, A. Pantea and D. Thomas (1998). “Chrysler Goes to "Extremes". Case Study.” Distributed Computing(October): 24-28.

Baskerville, R. (1998). Designing information systems security. Chichester, UK, John Wiley.

Baskerville, R., J. Travis and D. P. Truex (1992). Systems without method: The impact of new technologies on information systems development projects. Transactions on the impact of computer supported technologies in information systems development. K. E. Kendall, K. Lyytinen and J. I. DeGross. Amsterdam, Elsevier Science Publications: 241-260.

Bayer, S. and J. Highsmith (1994). “RADical software development.” American Programmer 7(6): 35-42.

Beck, K. (1999a). “Embracing Change With Extreme Programming.” IEEE Computer 32(10): 70-77.

Beck, K. (1999b). Extreme programming explained. Reading, Mass., Addison-Wesley.

Beck, K. (2000). Extreme Programming Explained: Embrace Change.

Beck, K., M. Beedle, A. Bennekum van, A. Cockburn, W. Cunningham, M. Fowler, J. Grenning, J. Highsmith, A. Hunt, R. Jeffries, J. Kern, B. Marick, R. Martin, S. Mellor, K. Schwaber, J. Sutherland and D. Thomas (2001). “Manifesto for Agile Software Development.” 2002(22.3.2002) http://AgileManifesto.org.

Bergquist, M. and J. Ljungberg (2001). “The power of gifts: organizing social relationships in open source communities.” Information Systems Journal 11(4): 305 - 320.

Boehm, B. (2002). “Get Ready For The Agile Methods, With Care.” Computer 35(1): 64-69.

Coad, P., E. LeFebvre and J. De Luca (2000). Java Modeling In Color With UML: Enterprise Components and Process, Prentice Hall.

Cockburn, A. (1998). Surviving Object-Oriented Projects: A Manager's Guide, Addison Wesley Longman.

Cockburn, A. (2000). Writing Effective Use Cases, The Crystal Collection for Software Professionals, Addison-Wesley Professional.

Cockburn, A. (2002a). Agile Software Development. Boston, Addison-Wesley.

Cockburn, A. (2002b). "Agile Software Development Joins the "Would-Be" Crowd." Cutter IT Journal 15(1): 6-12.

Coplien, J. O. (1998). A Generative Development Process Pattern Language. The Patterns Handbook. L. Rising. New York, Cambridge University Press: 243-300.

Coyne, R. (1995). Designing Information Technology in the Postmodern Age. Cambridge, Mass., MIT Press.

Crowston, K. and B. Scozzi (2002). "Open source software projects as virtual organisations: competency rallying for software development." IEE Proceedings - Software 149(1): 3 - 17.

Cunningham, W. (1996). Episodes: A Pattern Language of Competitive Development. Pattern Languages of Program Design 2. J. Vlissides. New York, Addison-Wesley.

DeMarco, T. and T. Lister (1999). Peopleware. New York, Dorset House.

Dempsey, B., D. Weiss, P. Jones and J. Greenberg (2002). "Who Is and Open Source Software Developer." CACM 45(2): 67 - 72.

Dhillon, G. (1997). Managing information systems security. London, UK, MacMillan Press.

Dobrica, L. and E. Niemelä (2002). "A survey on software architecture analysis methods." IEEE Transactions on Software Engineering 28(7): 638-653.

DSDMConsortium (1997). Dynamic Systems Development Method, version 3. Ashford, Eng., DSDM Consortium.

eWorkshop (2002). "Summary of the second eworkshop on agile methods." 2002(August 8) http://fc-md.umd.edu/projects/Agile/SecondeWorkshop/summary2ndeWorksh.htm.

Favaro, J. (2002). "Managing Requirements for Business Value." IEEE Software March/April: 15 - 17.

Feller, J. and B. Fitzgerald (2000). A Framework Analysis of the Open Source Software Development Paradigm. 21st Annual International Conference on Information Systems, Brisbane, Australia.

Gallivan, M. (2001). "Striking a balance between trust and control in a virtual organization: a content analysis of open source software case studies." Information Systems Journal 11(4): 273 - 276.

Ghosh, R. and V. Prakash (2000). "The Orbiten Free Software Survey." 2002(03052002) orbiten.org/ofss/01.html.

Gilb, T. (1988). Principles of Software Engineering Management. Wokingham, UK, Addison-Wesley.

Glass, R. L. (1999). "The realities of software technology payoffs." Communications of the ACM 42(2): 74-79.

Glass, R. L. (2001). "Agile Versus Traditional: Make Love, Not War!" Cutter IT Journal 14(12): 12-18.

Grenning, J. (2001). "Launching XP at a Process-Intensive Company." IEEE Software 18: 3-9.

Haungs, J. (2001). "Pair programming on the C3 project." Computer 34(2): 118-119.

Hawrysh, S. and J. Ruprecht (2000). Light Methodologies: It's Like Déjà Vu All Over Again. Cutter IT Journal. 13: 4 - 12.

Highsmith, J. (2002). Agile software development ecosystems. Boston, MA., Pearson Education.

Highsmith, J. and A. Cockburn (2001). “Agile Software Development: The Business of Innovation.” Computer 34(9): 120-122.

Highsmith, J. A. (2000). Adaptive Software Development: A Collaborative Approach to Managing Complex Systems. New York, NY, Dorset House Publishing.

Hightower, R. and N. Lesiecki (2002). Java Tools for Extreme Programming. New York, Wiley Computer Publishing.

Humphrey, W. S. (1995). A discipline for software engineering, Addison Wesley Longman, Inc.

Hunt, A., Thomas, D. (2000). The Pragmatic Programmer, Addison Wesley.

Jacobsen, I. (1994). Object-Oriented Software Engineering. New York, Addison-Wesley.

Jacobsen, I., M. Christerson, P. Jonsson and G. Overgaard (1994). Object-Oriented Software Engineering: A Use-Case-Driven Approach. Reading, MA, Addison-Wesley.

Jeffries, R., A. Anderson and H. C. (2001). Extreme Programming Installed. Upper Saddle River, NJ, Addison-Wesley.

Kruchten, P. (1996). “A Rational Development Process.” Crosstalk 9(7): 11-16.

Kruchten, P. (2000). The Rational Unified Process: an Introduction, Addison-Wesley.

Krueger, C. (2002). “Eliminating the adoption barrier.” IEEE Software 19(4): 29-31.

Lakoff, G. and M. Johnson (1998). Philosophy in the Flesh. New York, Basic Books.

Lerner, J. and J. Tirole (2001). “The Simple Economics of Open Source.” 2002(20.052002) http://www.people.hbs.edu/jlerner/publications.html.

Martin, R. C. (1998). The Process. Object Oriented Analysis and Design with Applications, Addison Wesley: 93-108.

Maurer, F. and S. Martel (2002a). “Extreme programming: Rapid development for Web-based applications.” IEEE Internet Computing 6(1): 86-90.

Maurer, F. and S. Martel (2002b). “On the Productivity of Agile Software Practices: An Industrial Case Study.” (page accessed August 16, 2002) http://sern.ucalgary.ca/~milos/papers/2002/MaurerMartel2002c.pdf.

McCauley, R. (2001). “Agile Development Methods Poised to Upset Status Quo.” SIGCSE Bulletin 33(4): 14 - 15.

Miller, D. and J. Lee (2001). “The people make the process: commitment to employees, decision making, and performance.” Journal of Management 27: 163-189.

Miller, G. G. (2001). The Characteristics of Agile Software Processes. The 39th International Conference of Object-Oriented Languages and Systems (TOOLS 39), Santa Barbara, CA.

Mockus, A., R. Fielding and J. Herbsleb (2000). A Case Study of Open Source Software Development: The Apache Server. 22nd International Conference on Software Engineering, ICSE 2000, Limerick, Ireland.

Moore, G. A. (1995). Inside the tornado. New York, HarperBusiness.

Nandhakumar, J. and J. Avison (1999). “The fiction of methodological development: a field study of information systems development.” Information Technology & People 12(2): 176-191.

Naur, P. (1993). “Understanding Turing’s universal machine: Personal style in program description.” The Computer Journal 36(4): 351-372.

O’Reilly, T. (1999). “Lessons from Open Source Software Development.” Communications of the ACM Vol. 42(No. 4): 32-37.

Palmer, S. R. and J. M. Felsing (2002). A Practical Guide to Feature-Driven Development.

Parnas, D. L. and P. C. Clements (1986). “A rational design process: How and why to fake it.” IEEE Transactions on Software Engineering 12(2): 251-257.

Pfeffer, J. (1998). The human equation : building profits by putting people first. Boston, MA, Harvard Business School Press, cop.

Pöyhönen, J. (2000). Uusi varallisuusoikeus. (In Finnish). Helsinki, Lakimiesliiton Kustannus.

Rising, L. and N. S. Janoff (2000). “The Scrum software development process for small teams.” IEEE Software 17(4): 26-32.

Schuh, P. (2001). “Recovery, Redemption, and Extreme Programming.” IEEE Software 18(6): 34-41.

Schwaber, K. (1995). Scrum Development Process. OOPSLA'95 Workshop on Business Object Design and Implementation, Springer-Verlag.

Schwaber, K. and M. Beedle (2002). Agile Software Development With Scrum. Upper Saddle River, NJ, Prentice-Hall.

Sharma, S., V. Sugumaran and B. Rajagopalan (2002). “A framework for creating hybrid-open source software communities.” Information Systems Journal 12(1): 7 - 25.

Smith, J. (2001). A Comparison of RUP and XP, Rational Software White Paper: http://www.rational.com/media/whitepapers/TP167.pdf.

Sol, H. G. (1983). A feature analysis of information systems design methodologies: Methodological considerations. Information systems design methodologies: A feature analysis. T. W. Olle, H. G. Sol and C. J. Tully. Amsterdam, Elsevier: 1-8.

Sommerville, I. (1996). Software engineering. New York, Addison-Wesley.

Song, X. and L. J. Osterweil (1991). Comparing design methodologies through process modeling. 1st International Conference on Software Process, Los Alamitos, Calif., IEEE CS Press.

Song, X. and L. J. Osterweil (1992). “Toward objective, systematic design-method comparisons.” IEEE Software 9(3): 43-53.

Stapleton, J. (1997). Dynamic systems development method - The method in practice, Addison Wesley.

Succi, G. and M. Marchesi (2000). Extreme Programming Examined: Selected Papers from the XP 2000 Conference. XP 2000 Conference, Cagliari, Italy, Addison-Wesley.

Sultan, F. and L. Chan (2000). “The adoption of new technology: The case of object-oriented computing in software companies.” IEEE Transactions on Engineering Management 47(1): 106-126.

Takeuchi, H. and I. Nonaka (1986). “The New Product Development Game.” Harvard Business Review Jan./Feb.: 137-146.

Truex, D. P., R. Baskerville and J. Travis (2000). “Amethodical systems development: The deferred meaning of systems development methods.” Accounting, Management and Information Technology 10: 53-79.

Wall, D. (2001). “using Open Source for a Profitable Startup.” Computer December: 158 - 160.

Warsta, J. (2001). Contracting in Software Business: Analysis of evolving contract processes and relationships. Department of Information Processing Science. Oulu, University of Oulu, Finland: 262.

Wiegers, K. E. (1998). “Read my lips: No new models.” IEEE Software 15(5): 10-13.

Williams, L., R. R. Kessler, W. Cunningham and R. Jeffries (2000). “Strengthening the Case for Pair Programming.” IEEE Software 17(4): 19-25.